\documentclass[12pt]{article}

\evensidemargin=\oddsidemargin
\usepackage{amsmath, amssymb, amsthm, mathrsfs, mathtools, amscd}
\usepackage{enumerate}
\usepackage{times}
\usepackage[normalem]{ulem}
\usepackage[nospace]{cite}
\usepackage{latexsym}
\usepackage{color,bm}
\usepackage{graphicx}
\usepackage{here}
\usepackage{hyperref} 
\usepackage{url}
\renewcommand{\path}{\textrm}

\usepackage{underscore}
\usepackage{dsfont}
\numberwithin{equation}{section}

\newtheorem{Theorem}{Theorem}[section]
\newtheorem{Proposition}[Theorem]{Proposition}    
 \newcommand{\bTheorem}{\begin{Theorem}}
\newcommand{\eTheorem}{\end{Theorem}}
\newcommand{\bProposition}{\begin{Proposition}}
\newcommand{\eProposition}{\end{Proposition}}
\newcommand{\bProof }{\begin{proof}}
\newcommand{\eProof}{\end{proof}}
\newcommand{\bitem}{\begin{itemize}\itemsep=0in}
\newcommand{\eitem}{\end{itemize}}
\newcommand{\benum}{\begin{enumerate}[{\rm (i)}]\itemsep=0in}
\newcommand{\eenum}{\end{enumerate}}
\newcommand{\deq}[1]{ \begin{align}#1\end{align}}
\newcommand{\deqs}[1]{ \begin{align*}#1\end{align*}}
\newcommand{\deqn}[1]{ \begin{align}#1\end{align}}
\newcommand{\deqed}[1]{\begin{equation}\begin{aligned}#1\end{aligned}\end{equation}}
\newcommand{\deql}[2]{\deqn{\label{eq:#1}{#2}}}
\newcommand{\cLL}{{\mathcal L}}
\newcommand{\cSS}{{\mathcal S}}

\newcommand{\N}{\mathds{N}}
\newcommand{\Q}{\mathds{Q}}
\newcommand{\R}{\mathds{R}}
\newcommand{\V}{V}

\newcommand{\p}{^\perp}
\newcommand{\al}{\alpha}

\newcommand{\be}{\beta}
\newcommand{\da}{\dagger}
\newcommand{\de}{\delta}

\newcommand{\et}{\eta}
\newcommand{\id}{\mathrm{id}}
\newcommand{\ie}{{\em i.e.}}

\newcommand{\la}{\lambda}
\newcommand{\mb}{\mbox}
\newcommand{\nn}{\nonumber}
\newcommand{\ph}{\phi}
\newcommand{\ps}{\psi}
\newcommand{\eq}[1]{(\ref{eq:#1})}

\newcommand{\ck}[1]{\check{#1}}
\newcommand{\cq}{\check{q}}
\newcommand{\cx}{\check{x}}
\newcommand{\cy}{\check{y}}
\newcommand{\cz}{\check{z}}

\newcommand{\De}{\Delta}
\newcommand{\Ga}{\Gamma}
\newcommand{\Or}{\vee}
\newcommand{\Ph}{\Phi}
\newcommand{\Ps}{\Psi}
\newcommand{\Eq}[1]{Eq.~(\ref{eq:#1})}
\newcommand{\VB}{V^{(\cB)}}
\newcommand{\VQ}{\V^{(\cQ)}}
\newcommand{\VR}{\V^{(\cR)}}
\newcommand{\Vtwo}{\V^{(\two)}}

\newcommand{\QH}{\cQ}
\newcommand{\RQ}{\R^{(\cQ)}}
\newcommand{\RQj}{\R^{(\cQ)}_{j}}

\newcommand{\LL}{\cL_{0}}
\newcommand{\bS}{\mathbf{S}}
\newcommand{\cA}{{\mathcal A}}
\newcommand{\cB}{{\mathcal B}}
\newcommand{\cC}{{\mathcal C}}
\newcommand{\cF}{{\mathcal F}}
\newcommand{\cH}{{\mathcal H}}
\newcommand{\cI}{{\mathcal I}}
\newcommand{\cL}{{\mathcal L}}
\newcommand{\cM}{{\mathcal M}}
\newcommand{\cN}{{\mathcal N}}
\newcommand{\cP}{{\mathcal P}}
\newcommand{\cQ}{{\mathcal Q}}
\newcommand{\cR}{{\mathcal R}}
\newcommand{\rB}{\mathrm{B}}
\newcommand{\rC}{\mathrm{C}}
\newcommand{\rR}{\mathrm{R}}
\newcommand{\rS}{\mathrm{S}}
\newcommand{\tP}{\tilde{P}}
\newcommand{\tQ}{\tilde{Q}}
\newcommand{\tX}{\tilde{X}}
\newcommand{\tY}{\tilde{Y}}
\newcommand{\Inf}{\bigwedge}
\renewcommand{\And}{\wedge}
\newcommand{\Andj}{*_j} 
 
\newcommand{\AndS}{*_\rS}
\newcommand{\AndC}{*_\rC}
\newcommand{\Implies}{\rightarrow}
\newcommand{\Iff}{\leftrightarrow}

\renewcommand{\and}{\wedge}
 
\newcommand{\astR}{*_\rR} 
\newcommand{\astS}{*_\rS}

\renewcommand{\iff}{\leftrightarrow}
\newcommand{\Not}{\neg}
\newcommand{\Then}{\rightarrow}  
\newcommand{\Thenj}{\rightarrow_j}
\newcommand{\ThenR}{\rightarrow_{\rR}}
\newcommand{\ThenS}{\rightarrow_{\rS}}
\newcommand{\ThenC}{\rightarrow_{\rC}}
\newcommand{\Sup}{\bigvee}
\newcommand{\dom}{\mathrm{dom}}

\newcommand{\bracket}[1]{\langle#1\rangle}

\newcommand{\rank}{{\rm rank}}
\newcommand{\remove}[1]{}
\newcommand{\two}{{\bf 2}}

\newcommand{\ckr}{\check{r}}
\newcommand{\iin}{\in\cR}
\newcommand{\val}[1]{[\![#1]\!]}
\newcommand{\vval}[1]{[\![#1]\!]_{j,\cQ}}
\newcommand{\valj}[1]{[\![#1]\!]_{j}}
\newcommand{\valB}[1]{\val{#1}_{\cB}}
\newcommand{\valQ}[1]{\val{#1}_{\cQ}}
\newcommand{\valS}[1]{\val{#1}_{\rS}}
\newcommand{\valC}[1]{\val{#1}_{\rC}}
\newcommand{\valR}[1]{\val{#1}_{\rR}}

\newcommand{\commutes}{\,\rotatebox[origin=c]{270}{$\multimap$}\,}
\newcommand{\cuniv}{\com}
\newcommand{\cdom}{\com}
\newcommand{\cm}{\com}
\newcommand{\com}{\mathrm{com}}
\newcommand{\tir}{\tilde{r}}
\newcommand{\tit}{\tilde{t}}
\newcommand{\tiu}{\tilde{u}}
\newcommand{\tiv}{\tilde{v}}
\newcommand{\tiw}{\tilde{w}}

\newcommand{\disp}{\displaystyle}

\renewcommand{\disp}{\displaystyle}
\newcommand{\bproof}{\begin{proof}}
\newcommand{\eproof}{\end{proof}}

\newcommand{\les}{\preccurlyeq}

\newcommand{\kQ}{\check{\Q}}
\newcommand{\kq}{\check{q}}
\newcommand{\SA}{\mathrm{SA}}

\newcommand{\dInf}{\displaystyle\Inf}
\newcommand{\dSup}{\displaystyle\Sup}
\newcommand{\ran}{\cR}

\title{\LARGE\bf Quantum set theory: quantum conditionals and order of observables\thanks{A preliminary account of this study was presented at the 13th International Conference on Quantum Physics and Logic (QPL 2016), University of Strathclyde, Glasgow, Scotland, June 6--10, 2016, and appeared in Ref.~\cite{17A1}.}}
\author{\sc Masanao Ozawa\thanks{Email: ozawa@is.nagoya-u.ac.jp $|$ ORCID: 0000-0002-9245-8187}
\\ \\
{\small\it Graduate School of Informatics, Nagoya University, Chikusa-ku, Nagoya 464-8601, Japan}\\
{\small\it Fundamental Quantum Science Program, TRIP Headquarters, iTHEMS}\\
{\small\it  RIKEN, Wako, Saitama 351-0198, Japan}\\
{\small\it Kinugasa Research Organization, Ritsumeikan University, Kyoto 603-8577, Japan}\\
{\small\it Center for Mathematical Science and Artificial Intelligence}\\
{\small\it Chubu University Academy of Emerging Sciences, Kasugai, Aichi 487-8501, Japan}
}
\date{}
\begin{document}
\maketitle
\sloppy
\begin{abstract}
A difficulty in quantum logic is the well-known arbitrariness in choosing a binary operation for conditional among three principal candidates called Sasaki, contrapositive Sasaki, and the relevance conditional, mainly chosen from syntactical grounds. A fundamental problem remains to clarify their semantical differences manifest in operational concepts in quantum theory. Here, we attempt such an analysis through quantum set theory, developing models of quantum set theory built upon quantum logics with those three conditionals, each of which defines different quantum logical truth-value assignment for set theoretical statements. We show that each of them satisfies the transfer principle to determine the truth values of theorems of the ZFC set theory and defines the internal reals bijectively corresponding to the observables of the quantum system under consideration. Then, the truth values of their equality relations are identical irrespective of the chosen conditionals. Interestingly, however, their order relations exhibit a strong dependence on the specific conditional employed, while the order relation attains full truth value if and only if Olson's spectral order relation holds. We further characterize the order relation in terms of experimentally accessible relations for outcomes of successive projective measurements of the corresponding observables, showing that each choice has its own operational meaning with symmetry between the Sasaki and the contrapositive Sasaki conditionals, in contrast to the majority view that favors the Sasaki conditional. Our findings reveal that quantum set theory yields empirically testable predictions concerning state-dependent binary relations between quantum observables, thereby extending Born’s probabilistic interpretation from propositions to relations.
\medskip

\newcommand{\sep}{, ~}
\medskip

\end{abstract}
\newpage
\tableofcontents
\newpage     
\section{Introduction}
While Birkhoff and von Neumann \cite{BvN36} introduced quantum logic as the
study of the ``propositional calculus'' for experimental propositions on a quantum system 
in 1936, it took a long time before Takeuti \cite{Ta81} introduced quantum set theory as 
a ``set theory based on quantum logic'' in 1981 to develop ``mathematics based 
on quantum logic.''  Unfortunately, the delay of the jump from a ``propositional calculus'' 
into a ``set theory'', associated with the names of Boole and Frege in the history of classical logic,  
has often brought about irrelevant criticisms against the field's immaturity.
Takeuti \cite{Ta81} showed that the internal real numbers in quantum set theory
are in one-to-one correspondence with the quantum observables of the corresponding quantum system.   
The expressive power of quantum set theory is strong enough not only to fully reconstruct
 quantum theory from the purely logical discipline, but also to extend 
the usual Born probabilistic interpretation of quantum theory from primitive, or atomic,
experimental propositions and their compound propositions 
to any {\em relations} such as equality and order, 
which in general cannot be expressed within the propositional calculus, 
between observables definable in the language of set theory \cite{14QST,16A2}.

In this paper, using the framework of quantum set theory,
we study a long-standing problem in quantum logic concerning 
 the arbitrariness in choosing a binary 
operation for conditionals \cite{Urq83}. 
Hardegree \cite{Har81} defined a material conditional 
on an orthomodular lattice as a polynomially definable binary operation 
satisfying three fundamental requirements and showed that there are 
exactly three binary operations satisfying those conditions: the Sasaki 
conditional, the contrapositive Sasaki conditional, and the relevance 
conditional. Naturally, a fundamental problem is to show how the form 
of the conditional follows from an analysis of the experimental concept in 
quantum theory testable by experiments \cite{Urq83}.  Here, we attempt such an 
analysis through quantum set theory. 

In this paper, we develop new interpretations of $\VQ$ with three quantum material conditionals 
that satisfy both all forms of De Morgan's law and the transfer principle introduced in \cite{21QTD};
see the Appendix for details. 
We show that the reals in the model and the truth values of their equality are the same for those three
conditionals.
Up to this point, interpretations with different conditionals have behaved indistinguishably. 
However, we reveal that the order relation between quantum reals crucially depends on the underlying conditionals.
We characterize the experimental meanings of those order relations,
which turn out to be closely related to the spectral order introduced by Olson \cite{Ols71}
playing a significant role in the topos approach to quantum theory \cite{DD14},
in terms of the joint probability of the outcomes of the successive projective 
measurements of two observables.
Those characterizations clarify their individual features 
and will play a fundamental role in future applications to quantum physics.

Quantum set theory cuts across two distinct branches of mathematics:
foundations of mathematics and foundations of quantum mechanics.
Aside from the Birkhoff--von Neumann quantum logic \cite{BvN36},
it originated from the methods of forcing introduced by Cohen \cite{Coh63,Coh64,Coh66}
for the independence proof of the continuum hypothesis.
After Cohen's work,  forcing subsequently turned out to be a central method in 
set theory and was also incorporated with various notions in mathematics, 
in particular,  the notion of sheaves \cite{FMS79,FS79}
and notions of sets in nonstandard logics
such as Boolean-valued models of set theory \cite{TZ73,Bel05},
by which Scott and Solovay \cite{SS67} reformulated the method of forcing,
topos \cite{Joh77},  and intuitionistic set theory \cite{Gra79}. 
Takeuti's quantum set theory \cite{Ta81} was a natural successor of those studies through 
his and his colleagues' attempts of Boolean-valued analysis that realized systematic
applications of the methods of forcing to problems in analysis \cite{Ta78,83BT,84CT}.

Birkhoff and von Neumann \cite{BvN36} argued that quantum logic is represented 
by a lattice of closed linear subspaces in the Hilbert space of state vectors of a quantum
system under consideration and that their logical operations correspond 
to the lattice operations of subspaces, 
whereas the logic of classical physics is represented by the Boolean algebra of 
Borel subsets of the phase space of the physical system modulo the sets of Lebesgue measure zero so that logical operations correspond to set operations in an obvious way.

The correspondence between classical mechanics and Boolean logic can be extended to its ultimate form using Boolean-valued analysis, introduced by Scott \cite{Sco69a} and developed by Takeuti 
\cite{Ta78,Ta79b,Ta79c,Ta83a,Ta83b,Ta88} and followers 
\cite{Eda83,Jec85,KK94,KK99,Nis91,83BT,83BH,84CT,85NC,85TN,86BT,90BB,94FN,95SB,Smi84}, as follows. 
Let $\VB$ be the Boolean-valued universe of set theory constructed from the complete Boolean algebra $\cB$ of the logic of classical mechanics described above.  Then there is a natural correspondence between the system's physical quantities 
and the internal real numbers of the model $\VB$.
By the ZFC transfer principle for Boolean-valued models, 
every theorem on real numbers provable in ZFC gives rise to a valid physical statement 
on physical quantities \cite{Ta78}, and this provides a purely logical reconstruction of 
classical mechanics by the Boolean-valued interpretation of ZFC set theory.

Quantum set theory aims at extending this logical reconstruction to quantum theory.
Following the formulation based on general von Neumann algebras given by the present author
\cite{07TPQ,14QST,16A2},
let $\cQ$ be the complete orthomodular lattice of projections in the von Neumann algebra $\cM$ of 
observables of a physical system $\bS$ \cite{BR79,BR81,Ara99}.  To develop the quantum set theory 
based on the ``logic'' $\cQ$ of the system $\bS$,
we construct the universe $\VQ$  of quantum set theory in a way analogous to the construction of
the Boolean-valued universe $\VB$ from $\cB$.
Then there is a natural correspondence between the (external) physical quantities 
of the system and the (internal) real numbers 
in the universe $\VQ$.
Accordingly, the experimental propositions in the sense of Birkhoff and von Neumann \cite{BvN36} 
can be embedded in the language of quantum set theory with the
$\cQ$-valued truth value assignment based on the universe $\VQ$.
Moreover, we have a suitable form of the transfer principle from theorems of ZFC to a valid statement on the universe $\VQ$ \cite{07TPQ}.

When the von Neumann algebra $\cM$ is abelian, the corresponding physical system is considered as a classical system, 
and the projection lattice $\cQ$ is a complete Boolean algebra, so $\VQ$ is nothing but a Boolean-valued universe
of set theory.
Takeuti's original quantum set theory \cite{Ta81} corresponds to the case where the von Neumann  algebra 
is a type I factor, or equivalently, the algebra of all bounded operators, on a Hilbert space $\cH$
of the state vectors of a quantum system.

Quantum set theory was further developed by the present author \cite{17A1,17A2,21A6,21QTD},
for the logic represented by arbitrary complete orthomodular lattices,  
including all complete Boolean algebras to unify with Boolean-valued models
of set theory, with the transfer principle for determining the quantum truth values of theorems
of the ZFC set theory.  
Takeuti's original truth value assignment fails to satisfy De Morgan's laws for bounded quantifiers, 
but recently, we found new formulations that improve the truth value assignment
so that they satisfy both De Morgan's law and the transfer principle \cite{21QTD}.

The relationship with another logical approach to quantum theory, the so-called topos quantum theory 
\cite{Ish97,Ish98,Doe11,Doe12}, was also studied.
We have unified both approaches in an appropriate framework 
of paraconsistent set theory \cite{Eva15,21BBQ}.

In Sections 2 and 3.1, we collect preliminaries on quantum logic based on a complete orthomodular
lattice $\cQ$ and quantum set theory based on the $\cQ$-valued universe $\VQ$. 
In Section 3.2,  we examine the properties of the equality relation in the universe $\VQ$ and show 
that the equality axioms do not hold in general. Specifically, the equality in $\VQ$ satisfies reflexivity 
and symmetry but does not satisfy transitivity nor substitution law.
In Section 3.3, we show that there is a broad class of quantum sets that satisfy the equality axioms. 
This class is called the first-order quantum sets, which include the internal reals. 
This suggests that the violation of the equality axioms in quantum set theory is not really a serious obstruction 
in our attempts to reconstruct quantum theory, in contrast to a prevailing criticism \cite{Gib87}.
In Section 3.4, we define an internal real as an upper segment with the end point, if any,
of a Dedekind cut of the internal rational numbers. 
In Section 3.5, we study commutators and the equality relation among them.
We show that the truth value of the equality relation between internal reals 
is not affected by the choice of conditionals.  We also show that 
the commutativity for internal reals follows from the equality.
In Section 4, we introduce logics as projection lattices of von Neumann algebras,
and we show that there is a one-to-one correspondence, called the 
Takeuti correspondence, between internal reals in $\VQ$ and self-adjoint operators 
affiliated with the von Neumann algebra generated by $\cQ$.
In Section 5, we study the order relation between internal reals.
We show that the relation holds for two internal reals with truth value 1 
if and only if the spectral order introduced by Olson \cite{Ols71} holds
for the corresponding self-adjoint operators irrespective of the choice of 
conditionals that determines the relevant truth value.
However, for the intermediate truth values, the relation crucially depends
on the choice of conditionals.
We study the experimental meanings of their truth values by comparing the probability of the 
relation determined by the Born rule with the probability of the order relation between the 
outcomes of successive projective measurements of the corresponding observables. We show that 
the choice of the conditional operation relates to the order of the measurements.
Section 10 concludes the paper with some concluding remarks.

\section{Quantum logic}
\subsection{Complete orthomodular lattices}
A {\em complete orthomodular lattice}  is a complete
lattice $\cQ$ with an {\em orthocomplementation},
a unary operation $\perp$ on $\cQ$ satisfying
(i)  if $P \le Q$, then $Q^{\perp}\le P^{\perp}$,
(ii) $P^{\perp\perp}=P$,
(iii) $P\Or P^{\perp}=1$ and $P\And P^{\perp}=0$,
where $0=\Inf\cQ$ and $1=\Sup\cQ$,
that satisfies the {\em orthomodular law}
(OM): if $P\le Q$, then $P\Or(P^{\perp}\And Q)=Q$.
In this paper, any complete orthomodular lattice is called a {\em logic}.
A non-empty subset of a logic $\cQ$ is called a {\em subalgebra} iff
it is closed under $\And $, $\Or$, and $\perp$.
A subalgebra $\cA$ of $\cQ$ is said to be {\em complete} iff 
any subset of $\cA$ has its supremum and infimum in $\cA$ that should coincide with originally defined in $\cQ$.
For any subset $\cA$ of $\cQ$, 
the subalgebra generated by $\cA$ is denoted by
$\Ga_0\cA$.
We refer the reader to Kalmbach \cite{Kal83} for a standard text on
orthomodular lattices.

We say that $P$ and $Q$ in a logic $\cQ$ {\em commute}, in  symbols
$P\commutes Q$, if  $P=(P\And Q)\Or(P\And
Q^{\perp})$.  A logic $\cQ$ is a Boolean
algebra if and only if $P\commutes Q$  for all $P,Q\in\cQ$ \cite[pp.~24--25]{Kal83}.
For any subset $\cA\subseteq\cQ$,
we denote by $\cA^{!}$ the {\em commutant} 
of $\cA$ in $\cQ$ \cite[p.~23]{Kal83}, \ie, 
\deq{
\cA^{!}=
\{P\in\cQ\mid P\commutes Q \mbox{ for all }
Q\in\cA\}.
}
Then $\cA^{!}$ is a complete subalgebra of $\cQ$.
The center of $\cQ$, denoted by $Z(\cQ)$, is the set of elements 
of $\cQ$ that commute with every element of $\cQ$, \ie, $Z(\cQ)=\cQ^{!}$. 
A {\em sublogic} of $\cQ$ is a subset $\cA$ of
$\cQ$ satisfying $\cA=\cA^{!!}$. 
For any subset $\cA\subseteq\cQ$, the smallest 
sublogic including $\cA$ is 
$\cA^{!!}$, called the  {\em sublogic generated by
$\cA$}.
Then it is easy to see that a subset 
 $\cA$ is a Boolean sublogic, or equivalently, 
 a distributive sublogic, if and only if 
$\cA=\cA^{!!}\subseteq\cA^{!}$.
The center of $\cA^{!!}$ is given by $Z(\cA^{!!})=\cA^{!}\cap\cA^{!!}$.

\subsection{Commutators}
\label{se:CIQL}
Let $\cA\subseteq\QH$.  
Takeuti \cite{Ta81} introduced the {\em commutator} of $\cA$, denoted by  $\com(\cA)$, as 
\deqn{
\com(\cA)=\dSup\{E\in\cA^{!}\mid \mb{$P\And E\commutes Q\And E$ for all $P,Q\in\cA$}\}.
}  
For $\cA=\{P_1,\ldots,P_n\}$, we write $\com(\cA)=\com(P_1,\ldots,P_n)$.
Then we have \cite{16A2}
\bTheorem\label{th:commutators}
For any $P,Q\in\QH$, any finite subsets $\cF\subseteq \QH$, and any subsets $\cA\subseteq \QH$, 
the following statements hold.
\benum
\item 
$\com(P,Q)=(P\And Q)\Or(P\And Q\p)\Or(P\p\And Q)\Or(P\p\And Q\p)$.
\item 
$\com(\cF)=\dSup_{\theta:\cF\to\{\id,\perp\}}\dInf_{P\in\cF}P^{\theta(P)}$,
 where $\{\id,\perp\}$ stands for the set consisting of the identity operation $\id$ and the orthocomplementation~$\perp$.
\item  
$\com(\cA)=\max\{E\in\cA^{!}\mid \mb{\rm $P\And E\commutes Q\And E$ for any $P,Q\in\cA$}\}$.
\item  
$\com(\cA)=\max\{E\in\cA^{!}\cap\cA^{!!}\mid \mb{\rm $P\And E\commutes Q\And E$ for any $P,Q\in\cA$}\}$.
\item  $\com(\cA)=\max\{E\in\cA^{!}\cap\cA^{!!}\mid [0,E]\subseteq\cA'\cap\cA^{!!}\}$.
\item 
$\com(\cA)=\dInf\{\com(P,Q)\mid P,Q\in\cA^{!!}\}$.
\eenum
\eTheorem

The following theorem clarifies the significance of commutators of a subset of a complete
orthomodular lattice \cite[Theorem 2.4]{16A2}.

\sloppy
\bTheorem[Decomposition Theorem]\label{th:decomposition}
Let $\cA$ be a subset of a logic $\QH$.
Then  the sublogic $\cA^{!!}$ generated by $\cA$ is isomorphic to the direct product 
of the complete Boolean algebra
$[0,\com(\cA)]_{\cA^{!!}}$ and the complete orthomodular lattice
$[0,\com(\cA)^\perp]_{\cA^{!!}}$ without any non-trivial Boolean factors.
\eTheorem
We refer the reader to 
\cite{Ta81,Pul85,Che89,16A2} for further results about commutators
in orthomodular lattices.

\subsection{Material conditionals}
\label{se:GIIQL}
In classical logic, the conditional operation $\Then$ is defined by negation 
$\perp$ and
disjunction $\Or$ as $P\Then Q=P^{\perp}\Or Q$.
In quantum logic there is well-known arbitrariness in choosing a binary operation for conditional.
Following Hardegree \cite{Har81}, we define a {\em quantum material conditional} 
on a logic $\cQ$ as a binary operation $\Then$ on $\cQ$ definable 
by an ortholattice polynomial $p(x,y)$  as $P\Then Q=p(P,Q)$ for all $P,Q\in\cQ$ satisfying the following 
``minimum implicative conditions'':
\benum
\item[(E)]  $P\Then Q=1$ if and only if $P\le Q$. 
\item[(MP)] (\emph{modus ponens}) $P\And(P\Then Q)\le Q$.
\item[(MT)] (\emph{modus tollens}) $Q^{\perp}\And (P\Then Q) \le P^{\perp}$.
\eenum
 The following two theorems were found by Hardegree \cite{Har81},
 but only brute force proofs have been suggested.
 Analytic proofs have been given in  \cite[Theorem 3.8, Corollary 3.12]{21QTD}.
 \begin{Theorem}[Hardegree \cite{Har81}] Condition (E) implies the following condition: 
 \begin{itemize}
 \item[{\rm (LB)}] If $P\commutes Q$, then 
$P\Then Q=P^{\perp}\Or Q$ for all $P,Q\in\cQ$.
\end{itemize} 
 \end{Theorem}
 A binary ortholattice polynomial $\Then$ on a logic $\cQ$ is called a {\em quantized conditional} 
 iff it satisfies (LB).
 We discuss quantized logical operations further in the Appendix.

Hardegree \cite{Har81} showed that there are exactly three polynomially definable 
material conditionals as follows.
\begin{Theorem}[Hardegree \cite{Har81}]\label{th:poly-3}
There are exactly three polynomially definable material conditionals:
\benum
\item[{\rm (S)}] (Sasaki conditional) $P\ThenS Q: =P^{\perp}\Or(P\And Q)$.
\item[{\rm (C)}] (Contrapositive Sasaki conditional) $P\ThenC Q: =(P\Or Q)^{\perp}\Or Q$.
\item[{\rm (R)}] (Relevance conditional) $P\ThenR Q: =(P\And Q)\Or(P^{\perp}\And Q)\Or(P^{\perp}\And Q^{\perp})$.
\eenum
\end{Theorem}
We easily see the following relations among the three quantum
material conditionals.
 \begin{align}
P\ThenS Q&=P\ThenR Q\Or(P\p\And\com(P,Q)\p).\\
P\ThenC Q&=P\ThenR Q\Or(Q\And\com(P,Q)\p).\\
P\ThenS Q&=Q\p\ThenC P\p.\\
P\ThenC Q&=Q\p\ThenS P\p.\\
P\ThenR Q&=(P\ThenS Q)\And (P\ThenC Q). \label{eq:then-R}
\end{align}

We shall denote by $\Thenj$ with $j=\rS,\rC,\rR$ any one of the above material conditionals.
Once the conditional $\Then_{j}$ is specified, the associated biconditional  $\Iff_{j}$ is defined
by
\deq{
P\Iff_{j} Q = (P\Then_{j} Q)\And(Q\Then_{j} P).
}
Then it is easy to see that we have
\deq{
P\Iff_{\rS} Q = P\Iff_{\rC} Q= P\Iff_{\rR} Q=(P\And Q)\Or (P^{\perp}\And Q^{\perp}).
}
Thus, we write $\Iff$ for $\Iff_{j}$ for all $j=\rS,\rC,\rR$.
For the transitivity of conditionals and the biconditional, the following facts are known.
\bTheorem [Hardegree \cite{Har81}]\label{th:transitivity-biconditional}
 The following statements hold.
\benum
\item
For $j=\rS,\rC,\rR$, if $(P\Thenj Q)\And(Q\Thenj R)\le (P\Thenj R)$ holds for any $P,Q,R\in \cQ$
then $\cQ$ is a Boolean algebra.
\item
(Transitivity of biconditional) For any $P,Q,R\in \cQ$,   
\[(P\Iff Q)\And(Q\Iff R)\le (P\Iff R).\]
 \eenum
\eTheorem 
In a Boolean logic, the conditional and the conjunction are associated by
the relation $P\And Q=(P\Then Q^{\perp})\p$, and this relation plays
an essential role in the duality between bounded universal quantification
$(\forall x\in A)\ph(x)$ and bounded existential quantification
$(\exists x\in A)\ph(x)$.  
In order to keep the duality in quantum set theory to be developed 
later, we  introduce the binary  operation $*_j$ dual to $\Thenj$, called
a {\em dual conjunction}, by
\deq{
P*_j Q=(P\Thenj Q\p)\p.
}
We have the following relations:
\begin{align}
P *_{\rS}Q&=(P\And Q)\Or (P\And \com(P,Q)\p),\\
P *_{\rC}Q&=(P\And Q)\Or( Q\And\com(P,Q)\p),\\
P *_{\rR}Q&=(P\And Q)\Or\com(P,Q)\p.
\end{align}

So far, we have no general agreement on the choice among the above,  although the majority view favors the Sasaki conditional \cite{Urq83}.
In this paper, we develop quantum set theory based on each of 
the above three conditionals and examine the structure of  the internal real numbers.
We shall show that the quantum truth values for the equality relations do not depend on the choice among them, but that those for the order relations 
crucially depend on them.

In the rest of the paper, we shall use the general index $j$ for $j=\rS,\rC,\rR$
unless stated otherwise. 

\section{Quantum set theory}
\label{se:UQ}
\subsection{Orthomodular-valued universe}
We denote by $\V$ the universe of the Zermelo-Fraenkel set theory
with the axiom of choice (ZFC).
Let $\LL(\in)$ be the language for first-order theory with equality consisting of binary relation symbols
$=$ for equality and $\in$ for membership; logical symbols $\Not$  for  negation,  $\And$ for conjunction, 
and $\Implies$ for conditionals; the universal quantifier $\forall x$; the bounded universal quantifier $\forall x\in y$;
and no constant symbols.
For any class $U\subseteq\V$, the language $\LL(\in,U)$ is the one obtained by adding a name 
for each element of $U$.
We take the symbols $\Or$ for disjunction,  
$\Iff$ for biconditionals, $\exists x$ for the existential quantifier,
 and $\exists x\in y$ for the bounded existential quantifier as derived
symbols by defining as
\benum
\item[(D1)] $\ph\Or\ps:=\Not(\Not \ph\And \Not\ps)$,
\item[(D2)] $\ph\Iff\ps:=(\ph\Then \ps) \And (\ps\Then \ph),$
\item[(D3)] $\exists x\in y\,\ph(x):=\Not(\forall x\in y\,\Not\ph(x)),$
\item[(D4)] $\exists x\ph(x):=\Not(\forall x\,\Not\ph(x)).$
\eenum

We also define the relation symbol $\subseteq$ for set inclusion by
\benum\setcounter{enumi}{4}
\item[(D5)]  $x\subseteq y:=(\forall z\in x)x\in y.$
\eenum

Let $\cQ$ be a logic. 
For each ordinal $ {\al}$, let
\deq{
\V_{\al}^{(\cQ)} = \{u|\ u:\dom(u)\to \cQ \mbox{ and }
(\exists \be<\al)\,
\dom(u) \subseteq V_{\be}^{(\cQ)}\}.
}
Here, we denote by $\dom(f)$ the domain of a function $f$.
By $f:D\to R$, we mean that $f$ is a function defined on a set $D$  with values in a set $R$.

The {\em $\cQ$-valued universe} $\VQ$ is defined
by 
\deq{
  \VQ= \bigcup _{{\al}{\in}\mbox{On}} V_{{\al}}^{(\cQ)},
}
where $\mbox{On}$ is the class of all ordinals. 
For every $u\in\VQ$, the rank of $u$, denoted by
$\rank(u)$,  is defined as the least $\al$ such that
$u\in \VQ_{\al+1}$.
It is easy to see that if $u\in\dom(v)$, then 
$\rank(u)<\rank(v)$.

For $u\in\VQ$, we define the {\em support} 
of $u$, denoted by $L(u)$, by transfinite recursion on the 
rank of $u$ by the relation
\deq{
L(u)=\left(\bigcup_{x\in\dom(u)}L(x)\right)\cup\{u(x)\mid x\in\dom(u)\}\cup\{0\}.
}
For $\cA\subseteq\VQ$, we write 
$L(\cA)=\bigcup_{u\in\cA}L(u)$, and
for $u_1,\ldots,u_n\in\VQ$, we write 
$L(u_1,\ldots,u_n)=L(\{u_1,\ldots,u_n\})$.

Then we obtain the following characterization of
subuniverses of $\VQ$ \cite[Proposition 4.2]{21QTD}. 

\bProposition\label{th:sublogic}
Let $\cR$ be a sublogic of a logic $\cQ$ and $\al$ an
ordinal. For any $u\in \VQ$, we have
$u\in\VR_{\al}$  if and only if
$u\in \VQ_{\al}$ and $L(u)\subseteq\cR$. 
In particular, $u\in\VR$ if and only if
$u\in\VQ$ and $L(u)\subseteq\cR$. 
Moreover,  for any $u\in\VQ$, its rank in $\VR$ 
is the same as its rank in $\VQ$.
\eProposition

An induction on rank argument leads to the following \cite{Bel05}.
\bTheorem[Induction Principle for $\VQ$]
For any formula $\ph(x)$,
\deq{
\forall u\in \VQ[\forall u'\in\dom(u)\, [\ph(u')\Then\ph(u)]]\Then \forall u\in\VQ\ph(u).
}
\eTheorem

Note that the implication symbol $\Then$ in the above statement refers to the classical implication in our mathematical language.

In what follows, the symbol $\Thenj$ for a binary operation on $\cQ$ generally denotes one of 
the Sasaki conditional $\ThenS$, the contrapositive Sasaki 
conditional $\ThenC$, and the relevance conditional $\ThenR$,
and the symbol $*_j$ denotes their dual conjunctions $*_\rS,*_\rC,*_\rR$, respectively.
For any $u,v\in\VQ$, the $\cQ$-valued truth values 
$\vval{u = v}$ and $\vval{u  \in v}$ of
atomic formulas $u=v$ and $u\in v$ are assigned
by the following rules recursive in rank \cite[\S 4.2]{21QTD}.
\begin{enumerate}[(i)]\itemsep=0in
\item[(R1)] $\vval{u = v}
= \dInf_{u' \in  \dom(u)}(u(u') \Thenj
\vval{u'  \in v})
\And \dInf_{v' \in   \dom(v)}(v(v') 
\Thenj \vval{v'  \in u})$.
\item[(R2)] $ \vval{u \in v} 
= \dSup_{v' \in \dom(v)} (v(v')*_j \vval{v'=u})$.
\end{enumerate}

To each statement $\ph$ of $\LL(\in,\VQ)$ 
we assign the $\cQ$-valued truth value $ \val{\ph}_{j,\cQ}$ by the following
rules \cite[\S 4.2]{21QTD}.
\begin{enumerate}[(i)]\itemsep=0in
\item[(R3)] $ \vval{\Not\ph} = \vval{\ph}^{\perp}$.
\item[(R4)] $ \vval{\ph_1\And\ph_2} 
= \vval{\ph_{1}} \And \vval{\ph_{2}}$.
\item[(R5)] $ \vval{\ph_1\rightarrow\ph_2} 
= \vval{\ph_{1}} \Thenj \vval{\ph_{2}}$.
\item[(R6)] $ \vval{(\forall x\in u)\, {\ph}(x)} 
= \dInf_{u'\in \dom(u)}
(u(u') \Thenj \vval{\ph(u')})$.
\item[(R7)] $ \vval{(\forall x)\, {\ph}(x)} 
= \dInf_{u\in \VQ}\vval{\ph(u)}$.
\end{enumerate}
We call the $\cQ$-valued truth value assignment determined by the above rules the {\em $j$-interpretation.}

By the definitions of the derived logical symbols, (D1)--(D4),  we have the following relations.
\begin{enumerate}[(i)]\itemsep=0in
\setcounter{enumi}{13}
\item[(T1)] $ \vval{\ph_1\Or\ph_2} 
= \vval{\ph_{1}} \Or \vval{\ph_{2}}$.
\item[(T2)] $ \vval{\ph_1\Iff\ph_2} 
= \vval{\ph_{1}} \Iff \vval{\ph_{2}}$.
\item[(T3)] $ \vval{(\exists x\in u)\, {\ph}(x)} 
= \dSup_{u'\in \dom(u)}(u(u') *_j \vval{\ph(u')})$.
\item[(T4)] $ \vval{(\exists x)\, {\ph}(x)} 
= \dSup_{u\in \VQ}\vval{\ph(u)}$.
\end{enumerate}

Note that according to the above, we have the following relations.
\begin{enumerate}[(i)]\itemsep=0in
\setcounter{enumi}{17}
\item[(T5)] $\vval{u = v}=\vval{\forall x\in u(x\in v)\And \forall x\in v(x\in u)}$,
\item[(T6)] $\vval{u  \in v}=\vval{\exists x\in v(x=u)}.$
\end{enumerate}
We also have the following relations satisfying De Morgan's laws \cite[\S 4.2]{21QTD}:
\begin{enumerate}[(i)]\itemsep=0in
\setcounter{enumi}{19}
\item[(M1)] $\vval{\Not(\ph_1\And\ph_2)} =\vval{\Not \ph_1\Or \Not \ph_2},$
\item[(M2)] $\vval{\Not(\ph_1\Or \ph_2)}=\vval{\Not \ph_1\And \Not \ph_2},$
\item[(M3)] $\vval{\Not(\forall x\in u\,\ph(x))}=\vval{\exists x\in u\,(\Not \ph(x))},$
\item[(M4)] $\vval{\Not(\exists x\in u\,\ph(x))}=\vval{\forall x\in u\,(\Not \ph(x))},$
\item[(M5)] $\vval{\Not(\forall x\,\ph(x))}=\vval{\exists x\,(\Not \ph(x))},$
\item[(M6)] $\vval{\Not(\exists x\,\ph(x))}=\vval{\forall x\,(\Not \ph(x))}.$
 \end{enumerate}
A formula in $\LL(\in)$ is called a {\em
$\De_{0}$-formula}  iff it has no unbounded quantifiers.
The following theorem holds \cite[Theorem 4.3]{21QTD}.

\sloppy
\begin{Theorem}[$\De_{0}$-Absoluteness Principle]
\label{th:Absoluteness}
\sloppy  
Let $\cR$ be a sublogic of a logic $\cQ$.
For any $\De_{0}$-formula 
${\ph} (x_{1},{\ldots}, x_{n}) $ 
of $\LL(\in)$ and $u_{1},{\ldots}, u_{n}\in \VR$, 
we have
\deq{
\val{\ph(u_{1},\ldots,u_{n})}_{j,\cR}=\vval{\ph(u_{1},\ldots,u_{n})}.
}
\end{Theorem}

Henceforth, 
for any $\De_{0}$-formula 
${\ph} (x_{1},{\ldots}, x_{n}) $
and $u_1,\ldots,u_n\in\VQ$,
we abbreviate $\val{\ph(u_{1},\ldots,u_{n})}_j=\val{\ph(u_{1},\ldots,u_{n})}_{j,\cR}=
\vval{\ph(u_{1},\ldots,u_{n})}$,
which is the common $\cR$-valued truth value for 
$u_{1},\ldots,u_{n}\in\VR\subseteq\VQ$.

The universe $\V$  can be embedded in
$V^{(\two)}\subseteq\VQ$, where $\two=\{0,1\}$,
by the following operation $\check{ }: v\mapsto\check{v}$ 
defined by the $\in$-recursion: 
for each $v\in\V$, $\check{v} = \{\check{u}|\ u\in v\} \times \{1\}$. 
Then applying Theorem \ref{th:Absoluteness} to $\cR=\two$,  we have the following 
\cite[Theorem 4.8]{21QTD}.

\begin{Theorem}[$\De_0$-Elementary Equivalence Principle]
\label{th:2.3.2}
\sloppy  
For any $\De_{0}$-formula 
${\ph} (x_{1},{\ldots}, x_{n}) $ 
of $\LL(\in)$ and $u_{1},{\ldots}, u_{n}\in V$,
we have $\valj{\ph(\check{u}_{1},\ldots,\check{u}_{n})}\in\two$
and that 
\[
\bracket{\V,\in}\models {\ph}(u_{1},{\ldots},u_{n})
\quad\mbox{if and only if}\quad
\valj{\ph(\check{u}_{1},\ldots,\check{u}_{n})}=1.
\]
\end{Theorem}

Let $\cA\subseteq\VQ$.  The {\em commutator
of $\cA$}, denoted by $\cm(\cA)$, is defined by 
\deq{\label{eq:com}
\cuniv(\cA)=\com (L(\cA)).
}
For any $u_1,\ldots,u_n\in\VQ$, we write
$\cuniv(u_1,\ldots,u_n)=\cuniv(\{u_1,\ldots,u_n\})$.

We have the following transfer principle for bounded theorems of ZFC
with respect to any material conditionals $\Thenj$ \cite[Theorem 4.15]{21QTD}.

\begin{Theorem}[Transfer Principle]
\label{th:TP}
For any $\De_{0}$-formula ${\ph} (x_{1},{\ldots}, x_{n})$ 
of $\LL(\in)$ and $u_{1},{\ldots}, u_{n}\in\VQ$, if 
${\ph} (x_{1},{\ldots}, x_{n})$ is provable in ZFC, then
we have
\deq{
\valj{\ph({u}_{1},\ldots,{u}_{n})}\ge \cuniv(u_{1},\ldots,u_{n}).
}
\end{Theorem}

\subsection{Equality}
In this section, we examine the properties of the equality relation
in the universe $\VQ$ under the $j$-interpretation 
and show that the equality axioms do not hold in general.
In particular, the equality in $\VQ$ satisfies reflexivity and symmetry but
does not satisfy transitivity or substitution laws.

\begin{Theorem}\label{th:equality}
For any $u,v \in\VQ$, the following relations hold.
\benum
\item
$\valj{u=v}=\valj{v=u}$.
\item  
$\valj{u=u}=1$.
\item
$u(x)\le \valj{x\in u}$ for any $x\in\dom(u)$.
\eenum
\end{Theorem}
\bProof 
Relation (i) is obvious from symmetry in the definition.
We shall prove relations (ii) and (iii) by transfinite induction on 
the rank of $u$.  The relations hold trivially if $u$ is of the lowest
rank.  
Let $u\in\VQ$. 
We assume that the relations hold for those with lower rank than $u$. 
Let $x\in\dom(u)$. 
By the induction hypothesis, we have $\valj{x=x}=1$, so  we have
\[
\valj{x\in u}=\dSup_{y\in\dom(u)}(u(y)*_j\valj{x=y})
\ge u(x)*_j\valj{x=x}=u(x).
\]
Thus, assertion (iii) holds for $u$.  
Then we have $(u(x)\Thenj\valj{x\in u})=1$ for all $x\in\dom(u)$,
and hence $\valj{u=u}=1$ follows.  Thus, relations (ii) and (iii) hold
by transfinite induction.
\eProof 

Let $u_1,\ldots,u_n\in\VQ$.  The $\cQ$-valued set of $u_1,\ldots,u_n\in\VQ$, 
denoted by
$\{u_1,\ldots,u_n\}_\cQ$, is defined by $\dom(\{u_1,\ldots,u_n\}_\cQ)=
\{u_1,\ldots,u_n\}$ and 
$\{u_1,\ldots,u_n\}_\cQ(u_k)=1$ for $k=1,\ldots,n$.
Then we have
 \deq{
 \valj{v\in\{u_1,\ldots,u_n\}_\cQ}=
\valj{v=u_1\Or\cdots \Or v=u_n}
}
for any $v\in\VQ$.
In fact,  we obtain
\begin{align*}
\valj{v\in\{u_1,\ldots,u_n\}_\cQ}
&=\dSup_{u\in\dom(\{u_1,\ldots,u_n\}_\cQ)}\{u_1,\ldots,u_n\}_\cQ(u)\Andj\valj{u=v}\\
&=\dSup_{k=1,\ldots,n}\{u_1,\ldots,u_n\}_\cQ(u_k)\Andj \valj{v=u_k}\\
&=\dSup_{k=1,\ldots,n}1 \Andj \valj{v=u_k}\\
&= \valj{v=u_1}\Or\cdots\Or\valj{v=u_n}\\
&=\valj{v=u_1\Or\cdots\Or v=u_n}.
\end{align*}
In particular, we have $\valj{u\in\{v\}_\cQ}=\valj{u=v}$ for any $u,v\in\VQ$.
We also have 
\deq{
\valj{\{u\}_\cQ=\{v\}_\cQ}=\valj{u=v},
}
which follows from
\begin{align*}
\valj{\{u\}_\cQ=\{v\}_\cQ}
&=
\dInf_{u'\in\dom(\{u\}_\cQ)}(\{u\}_\cQ(u')\Thenj \valj{u'\in\{v\}_\cQ})\\
&\qquad\And
\dInf_{v'\in\dom(\{v\}_\cQ)}(\{v\}_\cQ(v')\Thenj \valj{v'\in\{u\}_\cQ})\\
&=\valj{u\in\{v\}_\cQ}
\And
\valj{v\in\{u\}_\cQ}\\
&=\valj{u=v}.
\end{align*}

In Theorem \ref{th:equality}, we have shown that the equality $=$
satisfies reflexivity ($u=u$) and symmetry ($u=v\Then v=u$) in $\VQ$ .
From the transfer principle (Theorem \ref{th:TP}), the following  modifications of transitivity
and the substitution laws hold.
\begin{Theorem}\label{th:equality81}
Let  $\cQ$ be a logic.
For any $u,u',v,v',w\in \VQ$, we have the following.
\benum
\item
$\cuniv(u,v,u')\And
\valj{u=u'}\And\valj{u\in v}\le \valj{u'\in v}$.
\item
$\cuniv(u,v,u')\And\valj{u\in v}\And\valj{v=v'}
\le \valj{u\in v'}$.
\item
$\cuniv(u,v,w)\And\valj{u=v}\And\valj{v=w}
\le \valj{u=w}$.
\eenum
\end{Theorem}
\bProof 
Since $(x=y\And x\in z)\Then y\in z$ is a theorem of ZFC,
it follows from the transfer principle that
\deqs{
\cm(u,v,u')\le (\valj{u=u'}\And\valj{u\in v}\Thenj\valj{ u'\in v}). 
}
By taking the infimum with $\valj{u=u'}\And\valj{u\in v}$ on both sides
and applying the modus ponens (MP), we have
\deqs{\lefteqn{\cm(u,v,u')\And\valj{u=u'}\And\valj{u\in v}}\quad\\ 
&\le\valj{u=u'}\And\valj{u\in v}\And (\valj{u=u'}\And\valj{u\in v}\Thenj\valj{ u'\in v}) \\
&\le \valj{ u'\in v}.
}
 Thus, relation (i) follows. Assertions (ii) and (iii) are proved similarly.
\eProof 

Takeuti \cite{Ta81} gave counterexamples for the transitivity and the substitution 
laws for equality in $\VQ$ for the case $\cQ=\Q(\cH)$, the standard quantum logic on a Hilbert space $\cH$, 
with the $\cI(\Then_3,\ast_5)$-interpretation, where $\Then_3=\ThenS$ and $\ast_5=\And$  
(see  Appendix \ref{se:interpretations} for the definition of the  $\cI(\Then,\ast)$-interpretation),
which satisfies the transfer principle but does not satisfy De Morgan's law for bounded quantifiers.  
Here, we extend his consideration to our interpretations with the conditional $\Thenj$ and the dual conjunction
$\ast_j$, \ie, the $\cI(\Then_j,\ast_j)$-interpretation with $j=0,2,3$,  as follows.

\bTheorem\label{th:c-ex-eq}
For any non-distributive logic $\cQ$, there exist $u,v,w,\tiu,\tiv,\tiw\in\VQ$ that do not
satisfy the following relations for $j=\rS,\rC,\rR.$
\benum
\item $\valj{u=v}\And\valj{v=w}\le \valj{u=w}$.
\item $\valj{u=v}\And\valj{u\in \tiw}\le \valj{v\in \tiw}.$ 
\item $\valj{\tiu=\tiv}\And\valj{w\in\tiu}\le \valj{w\in\tiv}.$
\eenum
\eTheorem
\bProof 
Since $\cQ$ is not distributive, there exist $Q_1,Q_2\in\cQ$ such that $\cm(Q_1,Q_2)\p\ne 0$.
Let $E=\cm(Q_1,Q_2)\p$, $P_1=Q_1\And E$, and $P_2=Q_2\And E$. 
Then we have 
\benum
\item $P_1\ne 0$, $P_2\ne 0$, and $\cm(P_1,P_2)\p=E$, 
\item $P_1\And P_2=P_1\And P_2\p=P_1\p\And P_2=0$, and
$P_1\p\And P_2\p=E\p$, 
\item $P_1\Or P_2=E$, and
$P_1\Or P_2\p=P_1\p\Or P_2=P_1\p\Or P_2\p=1$, 
\item $P_1\p \Or E=1$, and
$P_2\p\Or E=1$.
\item $P_2 {*_\rS} P_1=P_2, P_2 {*_\rC} P_1=P_1$, and $P_2 {*_\rR} P_1=E$,
\eenum
Let $a_1,a_2\in\VQ$ be such that $\dom(a_1)=\dom(a_2)=\{\ck{0}\}$, $a_1(\ck{0})=1$,
and $a_2(\ck{0})=P_1$.  Then we have $\valj{a_1=a_2}=P_1$.
We define $u,v,w\in\VQ$ by 
\deqs{
&\dom(u)=\dom(v)=\dom(w)=\{a_1,a_2\},&\\
& u(a_1)=P_2, \quad u(a_2)=P_1\p, 
\quad v(a_1)=P_2,\quad v(a_2)=1,\quad w(a_1)=1,\quad w(a_2)=P_1.&
}
Then we have
\begin{align*}
\valj{a_1\in u}&=P_2\Or (P_1\p *_j P_1)=P_2,\\
\valj{a_2\in u}&=P_1\p\Or (P_2*_j P_1)=1,\\
\valj{a_1\in v}&=P_2\Or (1 *_j P_1)=E,\\
\valj{a_2\in v}&=1\Or (P_2 *_j P_1)=1,\\
\valj{a_1\in w}&=1\Or (P_1*_j P_1)=1,\\
\valj{a_2\in w}&=P_1\Or (1 *_j P_1)=P_1.
\end{align*}
Thus, we have
\begin{align*}
\valj{u=v}&=(v(a_1)\Thenj \valj{ a_1\in u})\And (u(a_1) \Thenj \valj{a_1\in v})\\
&=(P_2\Thenj P_2)\And (P_2\Thenj E) \\
&=1,\\
\valj{v=w}&=(v(a_2)\Thenj \valj{a_2\in w})\And(w(a_1)\Thenj \valj{a_1\in v})\\
&=(1\Thenj P_1)\And(1\Thenj E)\\
&=P_1\And E\\
&=P_1\\
\valj{u=w}
&=(u(a_2)\Then \valj{a_2\in w})\And (w(a_1)\Thenj \valj{a_1\in u})\\
&=(P_1\p \Thenj P_1)\And (1\Thenj P_2)\\
&=P_1\And P_2\\
&=0.
\end{align*}
Thus, $u$, $v$, and $w$ do not satisfy 
\deqs{
\valj{u=v}\And\valj{v=w}\le \valj{u=w}.
}
Let $\tiw=\{w\}_\cQ$.
Then $\valj{v\in \tiw}=\valj{v=w}$ and $\valj{u\in \tiw}=\valj{u=w}$
by
\deqs{
\valj{v\in \tiw}&=\Sup_{w'\in\dom(\tiw)}\tiw(w')*_j\valj{w'=v}
=\tiw(w)*_j\valj{w=v}=1\And \valj{w=v}\\
&=\valj{w=v}.
}
Thus, $u$, $v$, and $\tiw$ do not satisfy 
\deqs{
\valj{u=v}\And\valj{v\in\tiw}\le \valj{u\in\tiw}.
}
Let $\tiu=\{u\}_{\cQ}$ and $\tiv=\{v\}_\cQ$.
Then  $\valj{\tiu=\tiv}=\valj{u=v}$, $\valj{w\in\tiu}=\valj{w=u}$,
and $\valj{w\in\tiv}=\valj{w=v}$. 
Therefore, $\tiu$, $\tiv$, and $w$ do not satisfy 
\deqs{
\valj{\tiu=\tiv}\And\valj{w\in\tiu}\le \valj{w\in\tiv}.
}
\eProof 

\subsection{First-order quantum sets}
In the preceding section, we have shown that in quantum set theory,
the equality axioms, in particular, the transitivity law and the substitution
law, do not hold in general.
In this section, we show that there is a broad class of quantum sets, 
called first order quantum sets, 
including the internal reals,
for which the equality axioms hold. This suggests that the violation of the equality
axioms in quantum set theory is not really a serious obstruction to our attempts to reconstruct quantum
theory on its base,  in contrast to a prevailing criticism.
See,  e.g., Gibbins \cite[p.~159--160]{Gib87}
\begin{quotation}
[$\ldots$] it seems that the revisionist should consider 
what quantum logical mathematics would look like, and whether quantum theory 
in particular could be reconstructed on its base. [$\ldots$]
Gaisi Takeuti has shown that a quantum logical set theory is nonextensional -- and hence quite unlike
anything we want to countenance as a `set theory' [$\ldots$].

On the note of inevitable agnosticism we leave
quantum logical mathematics, still skeptical of securing a sensible foundation for revisionist quantum logic,
[$\ldots$]. 
\end{quotation}
Note that in the above, the violations of the transitivity law and the substitution law for equality 
(see, Theorem \ref{th:c-ex-eq} for the present formulation)
are wrongly called nonextensional; note that the extensionality axiom is satisfied according to the definition of equality.

\bTheorem\label{th:extension}
Let $\ph(x_1,\ldots,x_n)$ be a $\De_0$-formula in $\LL(\in)$.
Let $u_1,\ldots,u_n, \bar{u}_1,\ldots,\bar{u}_n\in\VQ$. 
Suppose that $\dom(\bar{u_k})\supseteq\dom(u_k)$, that 
$\bar{u}_k(x)=u_k(x)$ if $x\in\dom(u_k)$, and that  
$\bar{u}_k(x)=0$ if $x\in\dom(\bar{u}_k)\setminus\dom(u_k)$ for $k=1,\ldots,n$. 
Then for $j=\rS,\rC,\rR$, we have 
\deq{
\valj{\ph(\bar{u}_1,\ldots,\bar{u}_n)}
=\valj{\ph(u_1,\ldots,u_n)}.
}
\eTheorem
\bProof 
Let $v \in\VQ$, $u=u_1,\ldots,u_n$, and $\bar{u}=\bar{u}_1,\ldots,\bar{u}_n$.  
Note that $0*_jp=0\And p=0$ for all $p\in\cQ$, since
$p\commutes 0$.
Hence, we have
\deqs{
\valj{v\in \bar{u}}
&=\dSup_{x\in\dom(\bar{u})}(\bar{u}(x)*_j\valj{v=x})\\
&=\dSup_{x\in\dom(u)}(u(x)*_j\valj{v=x})\\
&=\valj{v\in u}.
}
Since $v$ was arbitrary, we have
$\valj{x\in \bar{u}}=\valj{x\in u}$ for all $x\in\VQ$.  
It follows that $\valj{x\in \bar{u}}=\valj{x\in u}$ for all $x\in\dom(v)$. 
Since $p\commutes 0$, we have $0\Thenj p=0^\perp\Or p=1$ for all $p\in\cQ$.
Hence, we have
\deqs{
\valj{\bar{u}=v}
&=
\dInf_{x\in\dom(\bar{u})}(\bar{u}(x)\Thenj\valj{x\in v})
\And
\dInf_{x\in\dom(v)}(v(x)\Thenj\valj{x\in \bar{u}})\\
&=
\dInf_{x\in\dom(u)}(u(x)\Thenj\valj{x\in v})
\And
\dInf_{x\in\dom(v)}(v(x)\Thenj\valj{x\in u})\\
&=
\valj{u=v}.
}
Thus, we have $\valj{\bar{u}=x}=\valj{u=x}$ for
all $x\in\VQ$.  It follows that $\valj{\bar{u}=x}=\valj{u=x}$
for all $x\in\dom(v)$.  Hence,  we have
\deqs{
\valj{\bar{u}\in v}
&=\dSup_{x\in\dom(v)}(v(x)*_j\valj{\bar{u}=x})\\
&=\dSup_{x\in\dom(v)}(v(x)*_j\valj{u=x})\\
&=\valj{u\in v}.
}
Thus, we have $\valj{\bar{u}\in x}=\valj{u\in x}$ for all $x\in\VQ$.
We have proved the assertion for atomic formulas,
and hence the assertion can be proved by induction on the length of the formula.
The only nontrivial steps are for the cases where $\ph(u):=(\forall x\in
u)\ph'(x,u)$.
In this case, using $0\Thenj p=1$ for all $p$, we have
\deqs{
\valj{(\forall x\in\bar{u})\ph'(x,\bar{u})}
&=
\dInf_{x\in\dom(\bar{u})}(\bar{u}(x)
\Thenj\valj{\ph'(x,\bar{u})})\\
&=
\dInf_{x\in\dom(u)}(u(x)\Thenj\valj{\ph'(x,u)})\\
&=
\valj{(\forall x\in u)\ph'(x,u)}.
}
The proof is completed. 
\eProof 

For any $X\in V$, define $\VQ_{1}(X)$ by
\deq{
\VQ_{1}(X)=\{u\in\VQ\mid \dom(u)=\dom(\check{X})\}
}
and define $\VQ_{1}(V)$ by
\deq{
\VQ_{1}(V)=\bigcup_{X\in V}\VQ_{1}(X).
}
A  quantum set $u\in\VQ$ is said to be
{\em of the first-order} if $u\in\VQ_{1}(V)$.

\begin{Theorem}\label{th:identity-A}
Let  $X\in V$ and 
$u,v\in\VQ_{1}(X)$. 
We have
\benum
\item
$\valj{\check{x}\in u}=u(\check{x})$ if $x\in X$.
\item 
$\valj{u\subseteq v}=
\dInf_{x\in X}(u(\ck{x})\Thenj v(\ck{x}))$.
\item
$\valj{u=v}=\dInf_{x\in X}(u(\ck{x})\Iff  v(\ck{x}))$.
\eenum
\end{Theorem}
\bProof 
(i) 
We have
\deqs{
\valj{\check{x}\in u}
&=\dSup_{u'\in\dom(u)}(u(u')*_j\valj{\check{x}=u'})
=\dSup_{u'\in\dom(\ck{X})}(u(u')*_j\valj{\check{x}=u'})\\
&=\dSup_{y\in X}(u(\check{y})*_j\valj{\check{x}=\check{y}})
=\dSup_{y\in X}(u(\check{y})\And\valj{\check{x}=\check{y}})
=u(\check{x}),
}
so assertion (i) follows.

(ii) 
From (i), we have
\deqs{
\valj{u\subseteq v}
&=
\dInf_{x\in X}(u(\ck{x})\Thenj\valj{\ck{x}\in v})
=
\dInf_{x\in X}(u(\ck{x})\Thenj v(\ck{x})).
}
Consequently, assertion (ii) follows.  

(iii) From (ii), we have
\deqs{
\valj{u=v}
&=\valj{u\subseteq v}\And\valj{v\subseteq u}\\
&=\dInf_{x\in X}(u(\ck{x})\Thenj v(\ck{x}))
\And \dInf_{x\in X}(v(\ck{x})\Thenj u(\ck{x}))\\
&=\dInf_{x,y\in X}[(u(\ck{x})\Thenj v(\ck{x}))\And (v(\ck{y})\Thenj u(\ck{y}))]\\
&\le \dInf_{x\in X}(u(\ck{x})\Iff v(\ck{x}))\\
&=\dInf_{x\in X}(u(\ck{x})\Thenj v(\ck{x}))
\And \dInf_{x\in X}(v(\ck{x})\Thenj u(\ck{x})).
}
Thus, assertion (iii) follows.
\eProof 

Now, we show that the equality axioms hold on $\VQ_{1}(V)$,

\begin{Theorem}\label{th:Equality-A}
For any $u,v,w\in \VQ_{1}(V)$
and $x,y\in V$,
the following relations hold.
\benum
\item
$\valj{\ck{x}=\ck{y}}\And\valj{\ck{x}\in v}\le \valj{\ck{y}\in v}$.
\item
$
\valj{\ck{x}\in u}\And\valj{u\subseteq v}\le \valj{\ck{x}\in v}.
$
\item
$
\valj{\ck{x}\in u}\And\valj{u=v}\le \valj{\ck{x}\in v}.
$

\item
$
\valj{u=v}\And\valj{v=w}\le \valj{u=w}.
$
\eenum
Moreover, if the following relation holds for any $u,v,w\in \VQ_{1}(V)$,
then $\cQ$ is a Boolean algebra.
\benum\setcounter{enumi}{4}
\item
$
\valj{u\subseteq v}\And\valj{v\subseteq w}\le \valj{u\subseteq w}.
$
\eenum
\end{Theorem}

\bProof 
(i) Relation (i) follows easily from $\valj{\ck{x}=\ck{y}}=\de_{x,y}$.

(ii)
Let $p=\valj{\ck{x}\in u}\And\valj{u\subseteq v}$.
From Theorem \ref{th:identity-A} (i), we assume $\ck{x}\in\dom(u)$ without any loss of generality. 
Then we have $p\le (u(\ck{x})\Thenj\valj{\ck{x}\in v})=(\valj{\ck{x}\in u}\Thenj\valj{\ck{x}\in v})$.
Thus, by modus ponens (MP), we have 
\deqs{
p\le \valj{\ck{x}\in u}\And  (\valj{\ck{x}\in u}\Then_j\valj{\ck{x}\in v})
\le \valj{\ck{x}\in v},
}
so relation (ii) holds.  

(iii) Relation (iii) follows from (ii) easily. 

(iv) By assumption, there is some $X\in V$ such that $\dom(u)\cup\dom(v)\cup\dom(w)\subseteq \dom(\ck{X})$.
From Theorem \ref{th:extension}, we assume $\dom(u)=\dom(v)=\dom(w)=\dom(\ck{X})$ without any loss of generality.
From  Theorem \ref{th:identity-A}, we have
\deqs{
\valj{u=v}\And\valj{v=w}
&=
\dInf_{x\in X}(\valj{\ck{x}\in u}\Iff\valj{\ck{x}\in v})
\And
\dInf_{x\in X}(\valj{\ck{x}\in v}\Iff\valj{\ck{x}\in w}) \\
&\le
\dInf_{x\in X}[(\valj{\ck{x}\in u}\Iff\valj{\ck{x}\in v})
\And
(\valj{\ck{x}\in v}\Iff\valj{\ck{x}\in w} )].
}
By the transitivity of biconditional (Theorem \ref{th:transitivity-biconditional} (ii)), we have
\[
 (\valj{\ck{x}\in u}\Iff\valj{\ck{x}\in v})
\And
(\valj{\ck{x}\in v}\Iff\valj{\ck{x}\in w} )
\le
 \valj{\ck{x}\in u}\Iff\valj{\ck{x}\in w} 
\]
for all $x\in X$.  Thus, we have 
\deqs{
\valj{u=v}\And\valj{v=w}\le
\dInf_{x\in X} (\valj{\ck{x}\in u}\Iff\valj{\ck{x}\in w} )
=\valj{u=w},
}
and relation (iv) follows.

(v) Let $P,Q,R\in\cQ$. Let $u,v,w\in \VQ_1(V)$ be such that $\dom(u)=\dom(v)=\dom(w)=\{\ck{0}\}$
and $u(\ck{0})=P$,  $v(\ck{0})=Q$, and  $w(\ck{0})=R$. Then $\valj{u \subseteq v}=u(\ck{0})\Thenj v(\ck{0})=
P\Thenj Q$.  Similarly, $\valj{u \subseteq v}=Q\Thenj R$ and $\valj{u \subseteq v}=Q\Thenj R$.
Thus, $\valj{u\subseteq v}\And\valj{v\subseteq w}\le \valj{u\subseteq w}$ if and only if
 $(P\Thenj Q)\And(Q\Thenj R)\le (P\Thenj R)$.  Since $P,Q$, and $R$ are arbitrary, the assertion on relation (v) follows
 from Theorem \ref{th:transitivity-biconditional} (i).
\eProof 

\subsection{Real numbers in quantum set theory}
\label{se:RN}
\renewcommand{\vval}[1]{[\![#1]\!]}

In ZFC, there is a predicate $\Q(x)$ meaning that $x$ is a rational number.
Denote by $\Q$ the set of rational numbers in ZFC, \ie, 
$\Q=\{x\mid \Q(x)\}$.
Let  $\cB$ be a sublogic of $\cQ$
and $\kQ\in V^{(\two)}$ the embedding of the set $\Q\in V$ in $V^{(\two)}\subseteq\VB\subseteq\VQ$. 
Then we have $\val{\Q(u)}_{\cB}=\val{u\in\kQ}_{\cB}$ for any  
$u\in\VB$ and $\val{\Q=\kQ}_{\cB}=1$\cite[p.~12]{Ta78}. 
Thus, we regard $\kQ$ as the set of rational numbers in $\VQ$.
However, the analogous relation $\val{\R=\ck{\R}}_{\cB}=1$ 
does not necessarily hold for real numbers \cite[p.~12]{Ta78}.
We define a real number in quantum set theory 
by a Dedekind cut of the rational numbers;
while there are many equivalent definitions in ZFC 
for a real number, they may not be equivalent 
 in quantum set theory.
More precisely, we identify
a real number with the upper segment of a Dedekind cut
such that the lower segment has no end point,
or equivalently, the complement of the lower segment without endpoint of a Dedekind cut.
The formal definition of  the predicate $\R(x)$, 
``$x$ is a real number,'' is expressed by
\deq{
\R(x)&:=
\forall y\in x(y\in\check{\Q})
\And \exists y\in\check{\Q}(y\in x)
\And \exists y\in\check{\Q}(y\not\in x)\nn\\
&  \And
\forall y\in\check{\Q}(y\in x\Iff\forall z\in\check{\Q}
(y<z \Then z\in x)).
}
Here, $<$ is the order relation between the standard rational numbers given as
$\val{\ck{u}<\ck{v}}=1$ if $u<v$ and $\val{\ck{u}<\ck{v}}=0$ otherwise for all $u,v\in\Q$. 
We define $\RQj$ for $j=\rS,\rC,\rR$ to be 
the external set of internal real numbers in $\VQ$ under the $\cI(\Then_j,\ast_j)$-interpretation
as follows.
\deq{
\RQj= \{u\in\VQ|\ \dom(u)=\dom(\check{\Q})
\ \mb{and }\valj{\R(u)}=1\}.
}
Note that $\RQj\subseteq \VQ_{1}(\Q)$.

\begin{Theorem}\label{th:identity-R}
Let $u,v\in\RQj$. 
We have
\benum
\item 
$\valj{\ckr\in u}=u(\ckr)$ for any $r\in\Q$.
\item
$\valj{u=v}=\dInf_{r\in \Q}(u(\ck{r})\Iff  v(\ck{r}))$.
\item $\valj{u= v}\in L(u,v)^{!!}$.
\eenum
\end{Theorem}
\bProof 
Since $\RQj\subseteq \VQ_{1}(\Q)$, assertion (i) and (ii) folllow from Theorem \ref{th:identity-A}.
Since $L(u,v)=\{u(\cx), v(\cy)\mid x,y\in \Q\}\cup\{0, 1\}$,
assertion (iii) follows easily from (ii).
\eProof 

The following theorem supports our assumption that $\RQj$ represents 
the external set of internal real numbers in $\VQ$. 

\begin{Theorem}\label{th:representation}
Let $u\in\VQ$. 
If $\cuniv(u)\And\valj{\R(u)}=1$, then there is a unique
$v\in\RQj$ such that $\cuniv(u,v)=1$ and 
$\valj{u=v}=1$.
\end{Theorem}
\bProof 
Since $\cuniv(u)=1$, there is a Boolean sublogic $\cB\subseteq \cQ$
such that $L(u)\subseteq L(u)^{!!}=\cB$.
From Proposition \ref{th:sublogic}, we have $u\in\VB$.
Let $v\in\VB$ be such that
$\dom(v)=\dom(\check{\Q})\subseteq \Vtwo$ and
$v(\check{x})=\valB{\check{x}\in u}\in\cB$ for all
$x\in\Q$.
Then $v\in \VB$, so $\cuniv(u,v)=1$, 
and $\val{\R(v)}_{\cB}=1$.
By the $\De_0$-Absoluteness Principle, we have
$\val{\R(v)}_{j}=1$.
By definition, we have $v(x)\Then\valj{x\in u}=1$
for all $x\in\dom(v)$, and hence  
$\valj{v\subseteq u}=1$.
By the $\De_0$-Absoluteness Principle, $\valB{u\subseteq\kQ}=1$, so we have
\deqs{
\valB{u\subseteq v}
&=
\valB{u\subseteq\kQ}\Then\valB{u\subseteq v}\\
&=
\valB{u\subseteq\kQ\Then  \forall x (x\in u \Then x\in v)}\\
&=\valB{\forall x((u\subseteq\kQ\And x\in u\And x\in\kQ) \Then x\in v)}\\
&=\valB{u\subseteq\kQ\Then\forall x(x\in\kQ\Then( x\in u\Then x\in v))}\\
&=\valB{u\subseteq\kQ}\Then\valB{\forall x(x\in\kQ\Then( x\in u\Then x\in v))}\\
&=\valB{u\subseteq\kQ}\Then\Inf_{q\in\dom(\kQ)}(\kQ(q)\Then\valB{ q\in u\Then q\in v})\\
&=\Inf_{r\in\Q}(\valB{\ck{r}\in u}\Then\valB{\ck{r}\in v}).
}
From Theorem \ref{th:identity-A},  $\valB{\ck{r}\in v}=v(\ck{r})=\valB{\ck{r}\in u}$, and hence
$\valB{\ck{r}\in u}\Then\valB{\ck{r}\in v}=1$ for all $r\in\Q$.  
Therefore, we obtain $\valB{u\subseteq v}=1$, so we have shown $\valB{u=v}=1$.
By the $\De_0$-Absoluteness Principle,  we conclude  $\valj{u=v}=1$.
To show the uniqueness, let $v_1,v_2\in\RQj$ be
such that $\valj{u=v_1}=\valj{u=v_2}=1$
and $\cuniv(u,v_1)=\cuniv(u,v_2)=1$.
Let $r\in\Q$. 
Then 
$\valj{\check{r}\in v_1}=
\valj{\check{r}\in u}$ follows from
$\valj{u=v_1}=\cuniv(u,v_1,\check{r})=1$,
and we have
$\valj{\check{r}\in v_2}=\valj{\check{r}\in u}$ similarly.
Thus,  we have 
$\valj{\check{r}\in v_1}=
\valj{\check{r}\in v_2}$.
Since $v_1(\check{r})=\valj{\check{r}\in v_{1}}$
and $v_2(\check{r})=\valj{\check{r}\in v_{2}}$,
the relation $\valj{v_1=v_2}=1$ follows easily.
\eProof

The internal set $\R_{j,\cQ}$ of internal real numbers in $\VQ$ under the $j$-interpretation
is defined by
\deq{
\R_{j,\cQ}=\RQj\times\{1\}.
}
From Theorem \ref{th:identity-R} (iii), the $\cQ$-valued truth value of equality 
between two internal reals is independent of the choice of conditionals for $j=\rS, \rC, \rR$.

\begin{Theorem}\label{th:q-real}
\begin{enumerate}[{\rm (i)}]\itemsep=0in
\label{th:RQ}
The following statements hold.
\item
For any $u\in\VQ$ with $\dom(u)=\dom(\check{\Q})$, we have
\deqs{
\displaystyle \vval{\R(u)}_j
=\dSup_{r\in\Q} u(\ck{r})
\And
\left(\dInf_{r\in\Q}u(\ck{r})\right)^\perp
\And
\dInf_{r\in\Q}\left(u(\ck{r})\Iff\dInf_{s\in\Q:r<s} u(\ck{s})\right).
}
\item
For any $u\in\VQ$ with $\dom(u)=\dom(\check{\Q})$, we have
$u\in\RQj$ if and only if it satisfies
\begin{enumerate}
\item[(a)] $\displaystyle\dSup_{r\in\Q} u(\ck{r})=1.$
\item[(b)] $\displaystyle\dInf_{r\in\Q}u(\ck{r})=0.$
\item[(c)] $u(\ck{r})=\displaystyle\dInf_{s\in\Q:r<s} u(\ck{s})$\quad for all \quad $r\in\Q$.
\eenum
\item 
$
\RQ_{\rS}=\RQ_{\rC}=\RQ_{\rR}.
$
\item
$\cuniv(u)=1$ for any $u\in\RQj$.
\eenum
\end{Theorem}
\bProof 
(i) Suppose $u\in\VQ$ and $\dom(u)=\dom(\check{\Q})$.   Then
\begin{align*}
\valj{\forall y\in u(y\in\ck{\Q})}
&=\dInf_{u'\in\dom(u)}(u(u')\Thenj \valj{u'\in\ck{\Q}})
=\dInf_{r\in \Q}(u(\ck{r})\Thenj \valj{\ck{r}\in\ck{\Q}})\\
&=\dInf_{r\in \Q}(u(\ck{r})\Thenj  1)=1.
\end{align*}
Thus, we have
\deqs{\valj{\R(u)}
&=
\valj{\exists y\in\ck{\Q}(y\in u)}\And
\valj{\exists y\in\ck{\Q}(y\not\in u)}\\
&\qquad\And
\valj{\forall y\in \ck{\Q}
[y\in u\Iff\forall z\in \ck{\Q} (y<z\Thenj z\in u)]}.}
We have 
\deqs{
\valj{\exists y\in\ck{\Q}(y\in u)}
&=\dSup_{r\in\Q}(\ck{\Q}(\ck{r})\Andj\valj{\ck{r}\in u})
=\dSup_{r\in\Q}(1\Andj\valj{\ck{r}\in u})
=\dSup_{r\in\Q}\valj{\ck{r}\in u}\\
&=\dSup_{r\in\Q}u(\ck{r}) ,\\
\valj{\exists y\in\ck{\Q}(y\not\in u)}
&=\dSup_{r\in\Q}(\ck{\Q}(\ck{r})\Andj\valj{\ck{r}\not\in u})
=\dSup_{r\in\Q}\valj{\ck{r}\in u}\p
=\dSup_{r\in\Q}u(\ck{r})\p\\
&=\left(\dInf_{r\in\Q}u(\ck{r})\right)^{\perp},
}
\vspace{-1em}
\begin{align*}
\lefteqn{\valj{\forall y\in \ck{\Q}(y\in u\Iff \forall z\in\check{\Q}(y<z \Thenj z\in u))}}\qquad\qquad\\
&={\dInf_{r\in\Q}(\ck{\Q}(\ck{r})\Thenj
\valj{\ck{r}\in u\Iff \forall z\in\ck{\Q}(\ck{r}<z \Thenj z\in u)})}\\
&={\dInf_{r\in\Q}(1\Thenj
\valj{\ck{r}\in u\Iff\forall z\in\ck{\Q}(\ck{r}<z \Thenj z\in u)})}\\
&={\dInf_{r\in\Q}
(\valj{\ck{r}\in u\Iff\forall z\in\ck{\Q}(\ck{r}<z \Thenj z\in u)})}\\
&={\dInf_{r\in\Q}
(\valj{\ck{r}\in u}\Iff\valj{\forall z\in\ck{\Q}(\ck{r}<z \Thenj z\in u)}})\\
&=\dInf_{r\in\Q}\left(\valj{\ck{r}\in u}\Iff
\dInf_{{s}\in\Q}(\Q(\ck{s})\Thenj\valj{\ck{r}< \ck{s} \Thenj \ck{s}\in u})\right)\\
&=\dInf_{r\in\Q}\left(\valj{\ck{r}\in u}\Iff
\dInf_{{s}\in\Q}(\valj{\ck{r}<\ck{s}} \Thenj \valj{\ck{s}\in u})\right)\\
&=\dInf_{r\in\Q}\left(u(\ck{r})\Iff\dInf_{{s}\in\Q:r<s}u(\ck{s})\right).
\end{align*}
Thus, assertion (i) holds.

(ii) Assertion follows easily from (i).

(iii)  Assertion holds, since  $\vval{\R(u)}_j$ is independent of the
choice of conditional from (i).

(iv) We have
\deqs{
L(u)
=\bigcup_{s\in\Q}L(\check{s})
\cup
\{u(\check{s})\mid s\in\Q\}
\cup\{0\}
=\{u(\check{s})\mid s\in\Q\}\cup\{0,1\},
}
so it suffices to show that each $u(\check{s})$
with $s\in\Q$ is mutually commuting. 
By definition, we have
\deqs{
\valj{\forall y\in\check{\Q}(y\in u\Iff\forall z\in\check{\Q}
(y<z \Then z\in u))}
=1.
}
Hence, we have
\deqs{
u(\check{s})=\dInf_{t\in\Q: s<t}u(\check{t}).
}
Thus, if $s_1<s_2$, then
$u(\check{s}_1)\le u(\check{s}_2)$,
so $u(\check{s}_1)\commutes u(\check{s}_2)$.
Thus, assertion (iv) holds, since $L(u)$ consists of mutually commutating elements of $\cQ$.
\eProof 

From Theorem \ref{th:identity-R} (iii)  and Theorem \ref{th:q-real} (ii)  above,
 we have shown that the set of internal real numbers and the $\cQ$-valued truth value of 
the equality relation are independent of the choice of conditionals $\Thenj$ for $j=\rS,\rC,\rR$;
henceforth, we shall write 
$\RQ:=\RQ_{\rS}=\RQ_{\rC}=\RQ_{\rR}$
$\val{u=v}:=\valS{u=v}=\valC{u=v}=\valR{u=v}$ for any $u,v\in\RQ$.

In what follows, we write $r\And s=\min\{r,s\}$
and $r\Or s=\max\{r,s\}$ for any $r,s\in\Q$.
Since $L(u,v)=\{u(\cx), v(\cy)\mid x,y\in\Q\}\cup\{0,1\}$,
it follows easily from Theorems \ref{th:commutators} (iii) and \ref{th:q-real} (iv)
that the commutator $\cuniv(u,v)$ of $u,v\in\RQ$ satisfies
\deql{commutator-uv}{
 \cm(u,v)=\dSup\{E\in L(u,v)^{!}\mid
 u(\check{x})\And E\commutes v(\check{y})\And E
\mbox{ \/ for all }x,y \in \Q\},
}
and from Theorem \ref{th:commutators} (iv), we also have
\deql{commutator-uvmax}{
 \cm(u,v)=\max\{E\in L(u,v)^{!}\And L(u,v)^{!!} \mid 
 u(\check{x})\And E\commutes v(\check{y})\And E
\mbox{ \/ for all }x,y \in \Q\},
}
from which we have $\com(u,v)\in L(u,v)^{!!}$.

The equality axioms hold in $\RQ$ as follows, and the general failure of the equality axiom
in $\VQ$ is not a serious obstruction to applying quantum set theory to analysis and quantum
theory.

\begin{Theorem}\label{th:Equality}
For any $u,v,w\in \RQ$
and $x,y\in \Q$,
the following relations hold.
\benum
\item
$\valj{\ck{x}=\ck{y}}\And\valj{\ck{x}\in v}\le \valj{\ck{y}\in v}$.
\item
$
\valj{\ck{x}\in u}\And\valj{u\subseteq v}\le \valj{\ck{x}\in v}.
$
\item
$
\valj{\ck{x}\in u}\And\valj{u=v}\le \valj{\ck{x}\in v}.
$

\item
$
\valj{u=v}\And\valj{v=w}\le \valj{u=w}.
$
\eenum
Moreover, if the following relation holds for any $u,v,w\in \RQ$,
then $\cQ$ is a Boolean algebra.
\benum\setcounter{enumi}{4}
\item
$
\valj{u\subseteq v}\And\valj{v\subseteq w}\le \valj{u\subseteq w}.
$
\eenum
\end{Theorem}

\bProof 
Since $\RQ\subseteq \VQ_1(\Q)$, the assertions follow from 
Theorem \ref{th:Equality-A}.
\eProof 

\begin{Theorem}\label{th:equality-axioms}
The following relations hold in $\VQ$.
\benum
\item
$\valj{(\forall u\in\R_\cQ) u=u}=1$.
\item
$\valj{(\forall u,v\in\R_\cQ) u=v\Then v=u}=1$.
\item
$\valj{(\forall u,v,w\in\R_\cQ) u=v\And v=w\Then u=w}=1$.
\item
$\valj{(\forall v\in\R_\cQ)(\forall x,y\in v)x=y\And x\in v\Then y\in v}=1$.
\item
$\valj{(\forall u,v\in\R_\cQ)(\forall x\in u)x\in u\And u=v\Then x\in
v}=1.$
\eenum
\end{Theorem}

\bProof 
Relations (i) and (ii) follow from Theorem \ref{th:identity-R} (iii).
Relation (iii) follows from Theorem \ref{th:Equality} (iv).
In fact, we have 
\deqs{\lefteqn{\valj{(\forall u,v,w\in\R_\cQ) u=v\And v=w\Then u=w}}\qquad\\
&=
\dInf_{u,v,w\in\dom(\R_\cQ)}
\R_\cQ(w)\!\Thenj\! (\R_\cQ(v)\!\Thenj\! (\R_\cQ(u)\!\Thenj\!\valj{u=v\And v=w\Then u=w}))\\
&=\dInf_{u,v,w\in\RQ}
1\Thenj (1\Thenj (1\Thenj \valj{u=v\And v=w\Then u=w}))\\
&=\dInf_{u,v,w\in\RQ}
\valj{u=v\And v=w\Then u=w}\\
&=\dInf_{u,v,w\in\RQ}
[(\valj{u=v}\And\valj{ v=w})\Thenj \valj{u=w}]\\
&=1.
}
Similarly, 
relations (iv) and (v) follow from \ref{th:Equality} (i) and (iii), respectively.
\eProof 

\section{Quantum set theory on von Neumann algebraic logic}
\subsection{von Neumann algebraic logic}
Let $\cH$ be a Hilbert space with inner product $(\cdots,\cdots)$.
For any subset $S\subseteq\cH$,
we denote by $S^{\perp}$ the orthogonal complement
of $S$.
Then $S^{\perp\perp}$ is the closed linear span of $S$.
Let $\cC(\cH)$ be the set of all closed linear subspaces in
$\cH$. 
With the set inclusion ordering, 
the set $\cC(\cH)$ is a complete
lattice. 
The operation $M\mapsto M^\perp$ 
is  an orthocomplementation
on the lattice $\cC(\cH)$, with which $\cC(\cH)$ is a logic, \ie,
a complete orthomodular lattice.

Denote by $\cB(\cH)$ the algebra of bounded linear
operators on $\cH$ with adjoint operation ${}^{\da}$,
 and by $\cQ(\cH)$ the set of projections on $\cH$, \ie,
 $\cQ(\cH)=\{P\in\cB(\cH)\mid P=P^{\da}=P^2\}$.
We define the {\em operator ordering} on $\cB(\cH)$ by
$A\le B$ if $(\ps,A\ps)\le (\ps,B\ps)$ for
all $\ps\in\cH$. 
For any $A\in\cB(\cH)$, denote by $\cR (A)\in\cC(\cH)$
the closure of the range of $A$, \ie, 
$\cR(A)=(A\cH)^{\perp\perp}$, and by $\cN(A)\in\cC(\cH)$
the kernel of $A$, \ie, $\cN(A)=\{\ps\in\cH\mid A\ps=0\}$,
which satisfy $\cN(A)=\cR(A^{\da})\p$.

For any $M\in\cC(\cH)$,
denote by $\cP (M)\in\cQ(\cH)$ the projection operator 
of $\cH$ onto $M$.
Then $\cR\cP (M)=M$ for all $M\in\cC(\cH)$
and $\cP\cR (P)=P$ for all $P\in\cQ(\cH)$,
and we have $P\le Q$ if and only if $\cR (P)\subseteq\cR (Q)$
for all $P,Q\in\cQ(\cH)$,
so $\cQ(\cH)$ with the operator ordering is also a logic
isomorphic to $\cC(\cH)$.
The lattice operations are characterized by 
$P\And Q={\mb{weak-lim}}_{n\to\infty}(PQ)^{n}$, 
$P^\perp=1-P$ for all $P,Q\in\cQ(\cH)$.

Let $\cA\subseteq\cB(\cH)$.
We denote by $\cA'$ the {\em commutant of 
$\cA$ in $\cB(\cH)$}.
A self-adjoint subalgebra $\cM$ of $\cB(\cH)$ is called a
{\em von Neumann algebra} on $\cH$ if 
$\cM''=\cM$.
For any self-adjoint subset $\cA\subseteq\cB(\cH)$,
$\cA''$ is the von Neumann algebra generated by $\cA$.
We denote by $\cQ(\cM)$ the set of projections in
a von Neumann algebra $\cM$.
For any $P,Q\in\cQ(\cH)$, we have 
$P\commutes Q$ if $[P,Q]=0$, where $[P,Q]=PQ-QP$.
For any subset $\cA\subseteq\cQ(\cH)$,
we denote by $\cA^{!}$ the {\em commutant} 
of $\cA$ in $\cQ(\cH)$.
For any subset $\cA\subseteq\cQ(\cH)$, the smallest 
sublogic including $\cA$ is the logic
$\cA^{!!}$ called the  {\em logic generated by $\cA$}.  
Then a subset $\cQ \subseteq\cQ(\cH)$ is a sublogic of $\cQ(\cH)$ if
and only if $\cQ=\cQ(\cM)$ for some von Neumann algebra
$\cM$ on $\cH$ \cite[Proposition 2.1]{07TPQ}.
Any sublogic of $\cQ(\cH)$ will be called a {\em (von Neumann algebraic) logic on $\cH$}. 

We have the following characterizations of conditionals in logics on Hilbert spaces.

\begin{Theorem}\label{th:equivalence-1}
Let $\cQ$ be a logic on a Hilbert space $\cH$.
Let $P,Q\in\cQ$.
Then we have the following relations.
\begin{enumerate}[{\rm (i)}]\itemsep=0in
\item $\cR(P\ThenS Q)=\{\ps\in\cH\mid Q^\perp P\ps=0\}=\cN( Q\p P)$.
\item $\cR(P\ThenC Q)=\{\ps\in\cH\mid PQ^\perp\ps=0\}=\cN(PQ\p)$.
\item $\cR(P\ThenR Q)=\{\ps\in\cH\mid Q^\perp P\ps=PQ^\perp\ps=0\}=\cN( Q\p P)\cap\cN(PQ\p)$.
\item $\cR(P{\Iff} Q)=\cN(P-Q)$.
\item $\cR(P{\ast}_\rS Q)=\cR(PQ)$.
\item $\cR(P{\ast}_\rC Q)=\cR(QP)$.
\item $\cR(P{\ast}_\rR Q)=\cR(PQ+QP)$.
\eenum
\end{Theorem}
\bProof 
To show (i) suppose $\ps\iin (P\ThenS Q)$.  Then we have
$\ps=P^{\perp}\ps+(P\And Q)\ps$,
so we have $Q^\perp P\ps=0$.
Conversely, suppose $Q^\perp P\ps=0$.
Then we have $P\ps=QP\ps+Q^{\perp}P\ps=QP\ps\in\cR(Q).$
Since $P\ps\in\cR(P)$, we have $P\ps\in\cR(P)\cap\cR(Q)=\cR(P\And Q)$.
It follows that $\ps=P^{\perp}\ps+P\ps=P^{\perp}\ps+(P\And Q)\ps\in\cR(P{\Then}_\rS Q)$. 
Thus, relation (i) holds.
Relation (ii) follows from the relation $P\ThenC Q=Q^{\perp}\ThenS P^{\perp}$.
Relation (iii) follows from the relation 
$P\ThenR Q=(P\ThenS Q)\And (P\ThenC Q)$.
To show relation (iv), suppose $\ps\in \cR(P{\Iff} Q)$.
Then $\ps\in\cR(P\ThenR Q)\cap\cR(Q\ThenR P)$, and hence
$ PQ^\perp\ps=0$ and $P^\perp Q\ps=0$, so $P\ps=PQ\ps=Q\ps$.  
Conversely, if $P\ps=Q\ps$, we have $Q^\perp P\ps=0$ and $P^\perp Q\ps=0$,
so $\ps\in\cR(P{\Iff} Q)$.  Thus, relation (iv) follows.
To show (v), note the relation $\cN(X)=\cR(X^\da)\p$ for any operator $X$. 
Thus, we have $\cR(P\astS Q)=\cR((P\ThenS Q\p)\p)=\cR(P\ThenS Q\p)\p
=\cN(QP)\p=\cR(PQ)$. Thus, relation (v) follows.  Relation (vi) holds similarly.
From \Eq{then-R}, $\cR(P\astR Q)=\cR([(P\ThenS Q\p)\And(P\ThenC Q\p)]\p)
=\cR[(P\AndS Q)\Or(P\AndC Q)]=\cR[\cR(PQ)\Or\cR(QP)]=\cR(PQ+QP)$.
Thus, relation (vii) follows.
\eProof 

We have the following characterizations of commutators 
 in logics on Hilbert spaces.
 
\begin{Theorem}\label{th:cuniv}
Let $\cQ$ be a logic on a Hilbert space $\cH$.
For any subset $\cA\subseteq\cQ$, we have
\[
\cdom(\cA)=\cP\{\ps\in\cH\mid
[P_{1},P_{2}]P_{3}\ps=0
\mb{ \textrm  for all }P_{1},P_{2}\in\cA \mb{ \textrm and }P_{3}\in\cA^{!!}\}.
\]
\end{Theorem}
\bProof 
From Theorem \ref{th:commutators} (iv), we assume $\cQ=\cQ(\cH)$ without
any loss of generality.   Let 
\deqs{F&=\max\{E\in\cA^{!}\mid  P_{1}\And E\commutes P_{2}\And E
\mb{ for all }P_{1},P_{2}\in\cA\},\\
G&=\cP\{\ps\in\cH\mid [P_{1},P_{2}]P_{3}\ps=0
\mb{ \textrm  for all }P_{1},P_{2}\in\cA \mb{ and }P_{3}\in\cA^{!!}\}.}
From Theorem \ref{th:commutators} (iii), $\com(\cA)=F$, and
we have to show $F=G$.
Let $\ps\iin (F)$, 
$P_{1},P_{2}\in\cA$, and $P_3\in\cA^{!!}$.
Since $F\in\cA^{!}$, we have $P_1\And F=P_1F$, $P_2\And F=P_2F$,
and $P_1P_2P_3F=(P_1F)(P_2F)(P_3F)=(P_2F)(P_1F)(P_3F)=P_2P_1P_3F$.
Thus, $[P_1,P_2]P_3F=0$. Since $F\ps=\ps$, we have $[P_1,P_2]P_3\ps=0$,
so $\ps\iin (G)$. It follows that $\cR(F)\subseteq\cR(G)$.
Conversely, suppose $\psi\iin (G)$.
Let $M=(\cA^{!!}\ps)^{\perp\perp}$, the closed linear subspace spanned by $\cA^{!!}\ps$.
Then $M$ is invariant under all
$P\in\cA$, so $\cP(M)\in\cA^{!}$.
Let 
\deq{N&=\{\xi\in\cH\mid [P_{1},P_{2}]\xi=0\mbox{ for
all }P_1,P_2\in\cA\}.}
  Then $N$
is a closed subspace such that $M\subseteq N$ by
assumption,
so $[P_{1},P_{2}]\cP(M)=0$.
Since $\cP(M)\in\cA^{!}$, we have $P_{1}\And \cP(M)
\commutes P_{2}\And \cP(M)$.  Thus, we have $\cP(M)\subseteq F$,
and $\ps\in M\subseteq \cR(F)$.
It follows that $\cR(G)\subseteq\cR(F)$.
Therefore, we have shown $\com(\cA)=F=G$.
\eProof 

\begin{Theorem}\label{th:com}
Let $\cQ$ be a logic on $\cH$
and let $\cA\subseteq\cQ$.
Then 
\deq{\label{eq:th=com}
\com(\cA)=\cP\{\ps\in\cH\mid [A,B]\ps=0 \mbox{ \textrm for all }A,B\in\cA^{!!}\}.}
\end{Theorem}
\bProof 
Let 
\deqs{G&=\cP\{\ps\in\cH\mid
[P_{1},P_{2}]P_{3}\ps=0
\mb{ \textrm  for all }P_{1},P_{2}\in\cA \mb{ and } P_{3}\in\cA''\},\\
H&=\cP\{\ps\in\cH\mid [A,B]\ps=0 \mb{ \textrm  for all }A,B\in\cA''\}.
}
From Theorem \ref{th:cuniv}, $\com(\cA)=G$, and it suffices to show
$G=H$.
Suppose $\ps\iin (H)$.
Let $P_{1},P_{2}\in\cA$ and $P_{3}\in\cA''$.
We have $P_{2}P_{3}\in\cA''$, and hence
$(P_{1}P_{2})P_{3}\ps=P_{1}(P_{2}P_{3})\ps=(P_{2}P_{3})P_{1}\ps=
P_{2}(P_{3}P_{1})\ps
=P_{2}(P_{1}P_{3})\ps
=(P_{2}P_{1})P_{3}\ps$.
It follows that $\ps\iin (G)$, so $\cR(H)\subseteq\cR(G)$.
Conversely, suppose $\ps\iin (G)$.
From Theorems \ref{th:cuniv} and \ref{th:commutators} (iv), we have 
\deqs{G=\com(\cA)=\max\{E\in\cA^{!}\cap\cA^{!!}
\mid \mb{$P\And E\commutes Q\And E$ for any $P,Q\in\cA$}\},}
and $G\in\cA^{!}\cap\cA^{!!}$.  Let $P,Q\in\cA$.
We have $[P,QG]=[PG,QG]=[P\And G,Q\And G]=0$.
Since $P\in\cA$ was arbitrary, we have $QG\in\cA'\cap\cA''$.
Since $Q\in\cA$ was arbitrary, we have $\cA''G\subseteq \cA'\cap\cA''$.
It follows that $ABG=BAG$ for any $A,B\in\cA''$.
Thus, if  $\ps\iin (G)$, then we have $AB\ps=BA\ps$
and $\ps\in\cR(H)$, so $\cR(H)=\cR(G)$.  Therefore, we conclude $G=H$.
\eProof 

\subsection{Takeuti correspondence}
\newcommand{\cOS}{{\mathcal O}_{\bS}}
Let $\cM$ be a von Neumann algebra on a Hilbert
space $\cH$, and let $\cQ=\cQ(\cM)$.
A linear mapping $X$ from a dense linear subspace $\dom(X)$ in $\cH$, called the {\em domain} of $X$,  into $\cH$ is called 
an {\em operator} in $\cH$.  An operator $X$ is called {\em closed} iff the graph $\{\ps\oplus X\ps\mid \ps\in\dom(X)\}$ is closed in $\cH\oplus\cH$. 
A closed operator $X$ in $\cH$ is said to be {\em affiliated} with $\cM$, in symbols $X\,\et\,\cM$, 
iff $U^{*}XU=X$ for any unitary operator $U\in\cM'$.
For any operator $X$ in $\cH$, we define its {\em adjoint} operator as an operator $X^{\da}$ in $\cH$
that satisfies 
\deq{\dom(X^{\da})&=\{\xi\in\cH\mid\exists \et\in\cH, \forall \ps\in\dom(X), (X\ps,\xi)=(\ps,\et)\},\\
(\ps,X^{\da}\xi)&= (X\ps,\xi) \quad\mb{ for all } \xi\in\dom(X^{\da}).}
An operator $X$ in $\cH$ is said to be {\em self-adjoint} iff $\dom(X^\da)=\dom(X)$ and $X^{\da}\xi=X\xi$ for all $\xi\in\dom(X)$.  All self-adjoint operators are closed.
 Denote by $\SA(\cH)$ the set of self-adjoint operators in  $\cH$.
A family $\{E(\la)\}_{\la\in\R}$ of projection operators on $\cH$ is called a {\em resolution of the identity} iff it satisfies the following conditions (S1)--(S3).

\benum
\item[(S1)] $\displaystyle\dInf_{\la\in\R}E(\la)=0$.
\item[(S2)] $\displaystyle\dSup_{\la\in\R}E(\la)=I$.
\item[(S3)] $\displaystyle\dInf_{\la\in\R: \la_0<\la}E(\la)=E(\la_0)$.
\eenum
From (S3), if $\la'\le\la''$, then $E(\la')\le E(\la'')$ and $\lim_{\la_0<\la, \la\to\la_0} (\xi,E(\la)\et)=(\xi,E(\la_0)\et)$ for all $\xi,\et\in\cH$.
Then any resolution of the identity  $\{E(\la)\}_{\la\in\R}$ defines a self-adjoint operator $X$ that satisfies the following conditions.
\deq{\dom(X)&=\Big\{\ps\in\cH\mid \int_{-\infty}^{+\infty}\la^2d(\|E(\la)\ps\|^2)<\infty\Big\},\\
(\phi,X\ps)&=\int_{-\infty}^{+\infty} \la \,d(\phi,E(\la)\ps)\}}
for any $\phi\in\cH,\ps\in\dom(X)$.
In this case, the resolution of the identity $\{E(\la)\}_{\la\in\R}$  is said to {\em belong} to the self-adjoint operator $X$.
Now, the spectral decomposition theorem for self-adjoint operators can be stated as follows \cite{vN32E}:
{\em For any self-adjoint operator $X$, there exists a unique resolution of the identity belonging to $X$.
In this case, $X$ is affiliated with a von Neumann algebra $\cM$ on $\cH$ if and only if $E(\la)\in\cM$
for all $\la\in\R$.}
In the following, the  resolution of the identity belonging to a self-adjoint operator $X$ is denoted by $\{E^{X}(\la)\}_{\la\in\R}$.

In the following theorem, we shall show that the internal reals in $\VQ$ are bijectively
represented by the self-adjoint operators affiliated with the von Neumann algebra $\cQ''$
generated by the logic $\cQ$. 

\bTheorem
\label{th:Takeuti-2}
Let $\cM$ be a von Neumann algebra on a Hilbert space $\cH$ such that $\cQ=\cQ(\cM)$
or $\cQ''=\cM$. 
The relations
\benum
\item $E^{X}(\la)=\displaystyle\dInf_{s\in\Q:\la<s} u(\ck{s})$ for any $\la\in\R$,
\item $u(\ck{r})=E^{X}(r)$ for any $r\in\Q$,
\eenum
 set up a one-to-one correspondence between $X\in\SA(\cM)$ and $u\in\RQ$.
\eTheorem
\begin{proof}
Let $u\in\RQ$.  For any $\la\in\R$, we define 
\deqs{E(\la)=\displaystyle\dInf_{s\in\Q:\la<s} u(\ck{s}).}
Then 
\deqs{
E(\la_0)=\dInf_{s\in\Q:\la_0<s} u(\ck{s})=\dInf_{\la\in\R:\la_0<\la}\Big(\dInf_{s\in\Q:\la<s} u(\ck{s})\Big)
=\dInf_{\la\in\R:\la_0<\la}E(\la),
}
so $E(\la)$ satisfies (S3), and it satisfies (S1)--(S2) similarly.  Thus, $\{E(\la)\}_{\la\in\R}$ is a 
resolution of the identity such that $E(\la)\in\cM$, and hence there is a self-adjoint operator $X\in\SA(\cM)$
such that $E(\la)=E^{X}(\la)$.  Thus relation (i) defines a mapping $\Ph:u\in\RQ\mapsto X\in\SA(\cM)$.
Conversely, suppose $X\in\SA(\cM)$ and $r\in\Q$.  Then  $E^{X}(r)\in\cQ$, and
we define $u\in\VQ$ by
\deqs{\dom(u)=\dom(\ck{\Q}), \quad u(\ck{r})=E^{X}(r)\quad\mb{ for all }r\in\Q.}
It follows from (S1)--(S3) that $u$ satisfies conditions (a)--(c) in Theorem \ref{th:q-real},
from which $u\in\RQ$.
Thus relation (ii) defines a mapping $\Ps:X\in\SA(\cM)\mapsto u\in\RQ$.
Now, we have
 \deqs{(\Ps\circ\Ph(u))(\ck{r})&=\Ps(X)(\ck{r})=E^{X}(r)=\dInf_{s\in\Q:r<s} u(\ck{s})=u(\ck{r}),\\
 E^{\Ph\circ\Ps(X)}(\la)&=E^{\Ph(u)}(\la)=\dInf_{s\in\Q:\la<s} u(\ck{s})= \dInf_{s\in\Q:\la<s} E^{X}(s)
 =E^{X}(\la)}
for any $r\in\Q$ and $\la\in\R$. Thus relations (i) and (ii) define a one-to-one correspondence
between $\RQ$  and $\SA(\cM)$. 
\end{proof}
We call the above correspondence between $\RQ$ and $\SA(\cM)$ 
the {\em Takeuti correspondence}, 
and we write $u=\tilde{X}$ and $X=\hat{u}$ for $u\in\RQ$ and $X\in\SA(\cM)$
satisfying relations (i) and (ii).

\subsection{Commutativity}
From now on, let $\cQ$ be a logic on a Hilbert space $\cH$ and let $\cM$ be
the von Neumann algebra generated by $\cQ$, \ie,  $\cM=\cQ''$;
or equivalently, let $\cM$ be a von Neumann algebra on $\cH$ and let $\cQ=\cQ(\cM)$.

Now, we show that the commutator $\cuniv(u,v)$ of $u,v\in\RQ$ is characterized as follows.

\begin{Theorem}\label{th:cuniv-pair}
For any $u,v\in\R^{(Q)}$, we have
\deq{\label{eq:cuniv-pair}
\cuniv(u,v)=\cP\{\ps\in\cH\mid
u(\cx)v(\cy)\ps=v(\cy)u(\cx)\ps \mb{ {\rm  for all }} x,y\in\Q\}.
}
\end{Theorem}
\bProof 
We have $L(u,v)=\{0,1,u(\cx), v(\cx)\mid x\in\Q\}$.
Let $E$ be the right-hand side of  \Eq{cuniv-pair}.
Suppose $\psi\in\cR(\cm(u,v))$.
Let $x,y,z\in\Q$.
Then 
\deqs{u(\cx)v(\cy)\psi&=u(\cx)v(\cy)\cm(u,v)\psi\\
&=u(\cx)\cm(u,v)v(\cy)\cm(u,v)\psi\\
&=[u(\cx)\And\cm(u,v)][v(\cy)\And\cm(u,v)]\psi\\
&=[v(\cy)\And\cm(u,v)][u(\cx)\And\cm(u,v)]\psi\\
&=v(\cy)u(\cx)\cm(u,v)\psi\\
&=v(\cy)u(\cx)\psi,
} 
and hence  $\psi\in \cR(E)$.
Thus, $\cR(\cm(u,v))\subseteq \cR(E)$, and we obtain $\cm(u,v)\le E$.
Conversely, suppose $\ps\in\cR(E)$.  
Then  we have 
\deqs{
u(\cx)v(\cy)u(\cz)\psi
&=u(\cx)u(\cz)v(\cy)\psi
=u(\cx\And\cz)v(\cy)\psi
=v(\cy)u(\cx\And\cz)\psi\\
&=v(\cy)u(\cx)u(\cz)\psi.
}
Consequently, $u(\cz)\psi\in\cR(E)$.
Similarly, $v(\cz)\psi\in\cR(E)$.
Since $z\in\Q$ was arbitrary, $\ran(E)$ is $L(u,v)$-invariant subspace,
so $E\in L(u,v)^{!}$.
From the definition of $E$, we have $u(\ck{x})v(\ck{y})E\ps=
v(\ck{y})u(\ck{x})E\ps$ for any $\ps\in\cH$, so 
$u(\ck{x})v(\ck{y})E=v(\ck{y})u(\ck{x})E$.  Since $E\in L(u,v)^{!}$,
we have $u(\ck{x})v(\ck{y})E=[u(\ck{x})\And E][v(\ck{y})\And E]$
and $v(\ck{y})u(\ck{x})E=[v(\ck{y})\And E][u(\ck{x})\And E]$, similarly.
From \Eq{commutator-uv}, we have $E\le \cm(u,v)$. 
Therefore, we conclude $\cm(u,v)=E$.
\eProof 

\subsection{Equality}

Let $\cQ$ be a logic on a Hilbert space $\cH$.
The $\cQ$-valued truth value of equality between two internal reals is characterized by the
vectors satisfying the corresponding equality.

\begin{Theorem}
\label{th:equality-R}
For any $u,v\in\R^{(\cQ)}$, we have
\deql{equality-R}{
 \valj{u=v}=\cP\{\ps\in\cH\mid u(\check{x})\ps=v(\check{x})\ps \mbox{ {\rm for all }}x\in \Q\}.
}
\end{Theorem}
\bProof 
From Theorem \ref{th:identity-R} (iii), we have
\deqs{
\valj{u=v}
&=
\dInf_{r\in \Q}(u(\check{r})\Iff v(\check{r})).
}
From Theorem \ref{th:equivalence-1} (iv), we have
\deqs{
u(\check{r})\Iff v(\check{r})
=
\cP\{\ps\in\cH\mid u(\check{r})\ps=
v(\check{r})\ps\}.
}
Thus, the assertion follows easily.
\eProof 

The $\cQ$-value of equality  $\valj{u=v}$ for $u,v\in\RQ$ is 
independent of the choice of the conditional and characterized
as follows.

\begin{Theorem}
\label{th:equality-R-2}
For any $u,v\in\R^{(\cQ)}$ we have
\deql{equality-R-2}{
 \valj{u=v}=\dSup\{E\in L(u,v)^{!}\cap L(u,v)^{!!}\mid
 u(\check{x})\And E=v(\check{x})\And E
\mbox{ \/ for all }x\in \Q\}.
}
\end{Theorem}
\bProof
Let $C$ be the right-hand side (RHS) of \Eq{equality-R} and $D$ the RHS of 
\Eq{equality-R-2}.
From Theorem \ref{th:equality-R}, it suffices to prove $C=D$.
From Theorems \ref{th:identity-R} (iii) and \ref{th:equality-R},  $C\in L(u,v)^{!!}$.
Let $\ps\in\cR(C)$ and $x,y\in\Q$.
Then we have
\deqs{u(\cx)u(\cy)\psi&=u(\cx\And \cy)\psi=v(\cx\And \cy)\psi=v(\cx)v(\cy)\psi=v(\cx)u(\cy)\psi,\\
u(\cx)v(\cy)\psi&=u(\cx)u(\cy)\psi=u(\cx\And \cy)\psi
=v(\cx\And \cy)\psi
=v(\cx)v(\cy)\psi.}
Since $x$ was arbitrary,  we have $u(\cy)\psi\in\cR(C)$ and $v(\cy)\psi\in\cR(C)$
for any $y\in\Q$.
Thus, $\cR(C)$ is $u(\cy)$-invariant and $v(\cy)$-invariant for all $y\in\Q$,
so $C\in L(u,v)^{!}$.  Since it is obvious that $u(\cx)\And C=v(\cx)\And C$
for all $x\in\Q$, it follows that $C \le D$.
Let $E\in L(u,v)^{!}\cap L(u,v)^{!!}$ be such that 
$u(\cx)\And E=v(\cx)\And E$ for all $x\in\Q$.  
Then it is obvious that $E\le C$, so we obtain $D\le C$.  Therefore, we conclude $C=D$.
\eProof

\begin{Theorem}
\label{th:equality-R-n}
For any $u_1,\ldots,u_n\in\R^{(\cQ)}$, we have the following relations.
\benum
\item $\valj{u_1=u_2\And\cdots\And u_{n-1}=u_n}=\disp \dInf_{k,l=1,\ldots,n}\valj{u_k=u_l}$.
\item $\valj{u_1=u_2\And\cdots\And u_{n-1}=u_n}$\\[2pt]
\qquad$= \cP\{\ps\in\cH\mid \mb{\rm $u_k(\cx)\ps=u_l(\cx)\ps$ for all $k,l=1,\ldots,n$ and $x\in\Q$}\}$.
\item $\valj{u_1=u_2\And\cdots\And u_{n-1}=u_n}\in L(u_1,u_2,\ldots,u_n)'$.
\eenum
 \end{Theorem}
\bProof 
From the transitivity of equality for internal reals (Theorem \ref{th:Equality} (v)) and 
Theorem \ref{th:equality-R}, we have  
\deqs{\lefteqn{\valj{u_1=u_2\And\cdots\And u_{n-1}=u_n}}\quad\\
 &=\dInf_{k=1,\ldots,n-1}\valj{u_k=u_{k+1}}\\
 &= \dInf_{k,l=1,\ldots,n}\valj{u_k=u_l}\\
 &=\dInf_{k,l=1,\ldots,n}\cP\{\ps\in\cH\mid u_k(\check{x})\ps=u_l(\check{x})\ps \mbox{ {\rm for all }}x\in \Q\}\\
 &=\cP\Big(\bigcap_{k,l=1,\ldots,n}\{\ps\in\cH\mid u_k(\check{x})\ps=u_l(\check{x})\ps \mbox{ {\rm for all }}x\in \Q\}\Big)\\
 &=\cP\{\ps\in\cH\mid \mb{\rm $u_k(\check{x})\ps=u_l(\check{x})\ps$ for all $k,l=1,\ldots,n$ and $x\in\Q$}\}.
 }
 Thus, (i) and (ii) hold.
 Let $E$ be the right-hand side of relation (ii).
 Let $k,l,m\in\{1,\ldots,n\}$, $x,y\in\Q$, and $\ps\in\cR(E)$.  Then we have
 \deqs{
 u_k(\cx)u_m(\cy)\ps
 &=u_k(\cx)u_k(\cy)\ps
 =u_k(\cx\And \cy)\ps
 =u_l(\cx\And \cy)\ps
 =u_l(\cx)u_l(\cy)\ps\\
&=u_l(\cx)u_m(\cy)\ps.
 }
Since $k,l,x$ were arbitrary, $u_m(\cy)\ps\in\cR(E)$, so $\cR(E)$ is $u_m(\cy)$-invariant.
Since $m,y$ were arbitrary, $\cR(E)$ is invariant under $L(u_1,u_2,\ldots,u_n)(=\{0,1,u_1(\cx),\ldots,u_n(\cx)\mid x\in\Q\})$.
 It follows that $E\in L(u_1,u_2,\ldots,u_n)'$.
 Thus, (iii) follows from (ii).
\eProof 

\subsection{Equality implies commutativity for internal reals}
The following theorems show that commutativity follows from equality in $\R^{(\cQ)}$.

\begin{Theorem}\label{th:eq-com-1}
For any $u,v\in\R^{(\cQ)}$, we have
\deqs{
\valj{u=v}\le \cuniv(u,v).
}
\end{Theorem}
\bProof 
Suppose $\ps\in\cR(\valj{u=v})$.
Let $x,y\in\Q$.
From Theorem \ref{th:equality-R},
\deqs{u(\check{x})v(\check{y})\ps=
u(\check{x})u(\check{y})\ps=
u(\check{x}\And\cy)\ps=
v(\check{x}\And\cy)\ps
=v(\check{y})v(\cx)\ps=v(\check{y})u(\cx)\ps.}
Thus, $\ps\in\cR(\cuniv(u,v))$ from Theorem \ref{th:cuniv-pair}.
Since $\ps$ was arbitrary, we obtain  $\cR(\valj{u=v})\subseteq\cR(\cuniv(u,v))$, 
and we conclude $\valj{u=v}\le \cuniv(u,v)$.
\eProof 

\begin{Theorem}\label{th:eq-com-2}
For any $u_1,\ldots,u_n\in\R^{(\cQ)}$, we have
\[
\valj{u_1=u_2\And\cdots\And u_{n-1}=u_n}\le 
\cuniv(u_1,\ldots,u_n).
\]
\end{Theorem}
\bProof 
Let $E=\valj{u_1=u_2\And\cdots\And u_{n-1}=u_n}$.
Let $k,l\in\{1,\ldots,n\}$ and $x,y\in\Q$.
Then from Theorem \ref{th:equality-R-n} (i), we have $E=\valj{u_k=u_l}\And E$, and from Theorem \ref{th:Equality} (iii), we obtain
\deqs{u_k(\cx)\And E=\valj{\cx\in u_k}\And\valj{u_k=u_l}\And E\le \valj{\cx\in u_l}\And E= u_l(\cx)\And E.}
Thus, we have $u_k(\cx)\And E= u_l(\cx)\And E$ by symmetry.
If $x\le y$, then $u_k(\cx)\And E\le u_l(\cy)\And E$, and if $x\ge y$, then $u_k(\cx)\And E\ge u_l(\cy)\And E$.
Thus, in any case, we have $u_k(\cx)\And E\commutes u_l(\cy)\And E$.
Since $E\in  L(u_1,\ldots,u_n)'$ from Theorem \ref{th:equality-R-n} (iii), the assertion follows from the
definition of $\cuniv(u_1,\ldots,u_n)$.
\eProof 

\section{Order relations on quantum reals}
\label{se:7}
\subsection{Orthomodular-valued models}
Let $u,v,w\in\RQ$.
The formula $v\subseteq u$ is a  $\De_0$-formula in $\cL(\in,\RQ)$ such that
\deqs{
\mbox{$v\subseteq u$}&:=\mb{$\forall x\in v(x\in u)$}.
}
The order of reals is defined through the inclusion relation of upper segments of Dedekind cuts,
so we introduce $\De_0$-formulas $u\le v$, $u< v$, and $u<v\le w$ in  $\cL(\in,\RQ)$ as
\deqs{
\mbox{$u\le v$}&:=\mbox{$v\subseteq u$},\\
\mbox{$u < v$}&:=\mbox{$(u\le v) \And \Not( u=v)$},\\
\mbox{$u<v\le w$}&:=\mbox{$(u<v) \And (v\le w)$}.
}

\bProposition\label{th:order}
The following relations hold.
\benum
\item
$\valj{u\le v}=
\dInf_{r\in \Q}(v(\ck{r})\Thenj u(\ck{r}))$ for any $u,v\in\RQ$.
\item $\valj{u\le v}\in L(u,v)^{!!}$ for any $u,v\in\RQ$.
\item
$\valj{(\forall x,y\in\R_\cQ)(x\le y \Iff (\forall z\in\kQ)(y\le z\Then x\le z))}=1.$
\item
$\valj{(\forall x,y\in\R_\cQ)(x = y \Iff (\forall z\in\kQ)(y\le z\Iff x\le z))}=1.$
\item $\valj{(\forall x,y\in\R_\cQ) x\le y\And y\le x\Then x=y}=1$.
\eenum
\eProposition
\bProof
(i) Since $\RQ\subseteq \VQ_{1}(\Q)$, assertion (i) follows from Theorem \ref{th:identity-A} (ii).

(ii) Since $L(u,v)=\{u (\cx), v(\cy)\mid x,y\in \Q\}\cup\{0, 1\}$,
assertion (ii) follows easily from (i).

(iii) Follows from (i).

(iv) Follows from Theorem \ref{th:identity-A} (iii).

(v) Follows from Theorem \ref{th:identity-A} (iii).
\eProof

Recall that for any $r\in \R$, the embedding $r
\in V\mapsto\check{r}\in\Vtwo\subseteq\VQ$ satisfies
\deqs{
\dom(\check{r})=\{\check{s}\mid s\in \Q,\ r\le s\}
\quad\mbox{and}\quad
\check{r}(\check{s})=1
}
for all $s\in\dom(\ckr)$, since $s\in r$ if and only if $r\le s$.
Obviously, we have $\com(\ckr)=1$, and 
from  the $\De_0$-elementary equivalence principle,
$\valj{\R(\ckr)}=1$.
Thus, according to Theorem \ref{th:representation}
there is a unique
$u\in\R^{(\cQ)}$ such that $\cuniv(u,\ckr)=1$ and 
$\valj{u=\ckr}=1$.
In order to explicitly determine the counterpart of $r\in\R$ in $\RQ$
found above, 
we define $\tilde{r}\in\RQ$ for any $r\in \R$ by
\deqs{
\dom(\tilde{r})=\dom(\check{\Q})\quad\mbox{and}\quad
\tilde{r}(\check{t})=\valj{\check{r}\le \check{t}}
}
for all $t\in\Q$. Then we have $\tir\in\RQ$, 
$L(\tilde{r})=\{0,1\}$,  
$\tilde{r}\in\V^{(\two)}$, and $\valj{\tir=\ckr}=1.$

\begin{Proposition}
Let $r\in\Q$, $s,t\in\R$, and $u\in\RQ$.
We have the following relations.
\benum
\item
$\valj{\check{r}\in\tilde{s}}=\valj{\check{s}\le \check{r}}.$

\item
$\valj{\tilde{s}\le\tilde{t}}=\valj{\check{s}\le\check{t}}.$

\item
${\displaystyle \valj{u\le\tilde{t}}=\dInf_{q\in\Q:t<q}u(\check{q})}.$
\eenum
\end{Proposition}

\bProof 
We have
\deqs{
\valj{\check{r}\in\tilde{s}}
&=
\dSup_{t\in\dom(\tilde{s})}(\tilde{s}(t)\ast_j\valj{t=\check{r}})
=
\dSup_{t\in\Q}(\tilde{s}(\check{t})\And\valj{\check{t}=\check{r}})
=
\tilde{s}(\check{r})
=
\valj{\check{s}\le \check{r}},
}
so (i) holds.  We have
\deqs{
\valj{\tilde{s}\le\tilde{t}}
&=
\valj{(\forall x\in\tilde{t})x\in\tilde{s}}
=
\dInf_{q\in\Q}(\tilde{t}(\check{q})\Then\valj{\check{q}\in\tilde{s}})
=
\dInf_{q\in\Q:t\le q}\valj{\check{s}\le \check{q}}
=
\valj{\check{s}\le \check{t}},
}
and hence (ii) holds.  We have
\deqs{
\valj{u\le\tilde{t}}
&=
\valj{(\forall x\in\tilde{t})x\in u}
=
\dInf_{q\in\Q}(\tilde{t}(\check{q})\Then\valj{\check{q}\in u})
=
\dInf_{q\in\Q}(\tilde{t}(\check{q})\Then u(\check{q}))\\
&=
\dInf_{q\in\Q:t\le q}u(\check{q}),
}
so (iii) holds.
\eProof 

\subsection{Von Neumann algebra-valued models}
Let $\cM$ be a von Neumann algebra on a Hilbert space $\cH$.
Denote by $\cQ=\cP(\cM)$  the lattice of projections in $\cM$.
Then $\cQ$ is a complete orthomodular lattice such that $\cQ^{''}=\cM$
with respect to the orthocomplementation.
Denote by $\SA(\cM)$ the set of self-adjoint operators affiliated with $\cM$.
We have a one-to-one correspondence, the Takeuti correspondence, 
between $u\in\RQj$ and $X\in \SA(\cM)$ such that
$u(\ck{r})=E^{X}(r)$ for all $r\in \Q$ and 
$E^{X}(\la)=\Inf_{r\in\Q:\la\le r} u(\ck{r})$ for all $\la\in\R$;
in this case, we write $u=\tX$ and $X=\hat{u}$.

Recall that for any $r\in\R$, the $\cQ$-valued set $\tir\in\RQ$ with $\dom(\tir)
=\dom(\kQ)$ is defined by
$\tir(\cq)=\valj{\ckr\le\kq}$ for all $q\in\Q$.
On the other hand, by the Takeuti correspondence, for any $X\in\SA(\cM)$,
the $\cQ$-valued set $\tX\in\RQ$ with $\dom(\tX)=\dom(\kQ)$
 is defined by 
$\tX(\cq)=E^{X}(q)$ for all $q\in\Q$.
Then note that for any $r\in\R$, we have the relation $\tir=(r1)\tilde{ }$, where $r1$ is
the scalar operator on $\cH$. In fact, we have 
$E^{r1}(q)=1$ if $r\le q$ and $E^{r1}(q)=0$ otherwise,
so $(r1)\tilde{ }(\cq)=E^{r1}(q)=\valj{\ckr\le\cq}=\tir(\cq)$.

Moreover, the following relations hold.

\begin{Proposition}\label{th:QBorel}
Let $r\in\Q$, 
$s,t\in\R$ and $X\in \SA(\cM)$.
For $j=\rR,\rC,\rS$, we have the following relations.
\benum
\item
$\valj{\check{r}\in\tilde{s}}=\valj{\check{s}\le \check{r}}
=E^{s1}(r)=E^{s1}((-\infty,r])$.
\item
$\valj{\tilde{X}\le\tilde{t}}=E^{X}(t)=E^{X}((-\infty,t])$.
\item
$\valj{\tilde{s}\le\tilde{t}}=\valj{\check{s}\le\check{t}}
=E^{s1}(t)=E^{s1}((-\infty,t])$.
\item
$\valj{\tilde{t}<\tilde{X}}=1-E^{X}(t)=E^{X}((t,\infty))$.
\item
$\valj{\tilde{s}<\tilde{X}\le \tilde{t}}=E^X({t})-E^X({s})=
E^{X}((s,t])$.
\item
$\valj{\tilde{X}< \tilde{t}}=\dSup_{r\in\Q:r<t} E^{X}(r)
=E^{X}((-\infty,t))$.
\item
$\valj{\tilde{t}\le \tilde{X}}=1-\dSup_{r\in\Q:r<t} E^{X}(r)
=E^{X}([t,+\infty))$.
\item
$\valj{\tilde{X}=\tilde{t}}
=E^{X}(t)-\dSup_{r\in\Q:r<t} E^{X}(r)=E^{X}(\{t\})$.
\eenum
\end{Proposition}
\bProof 
(i) From Theorem \ref{th:identity-R} (i), we have
\deqs{
\valj{\check{r}\in\tilde{s}}
&=
\tilde{s}(\check{r})=\valj{\check{s}\le \check{r}}=E^{s1}((-\infty,r]),
}
so (i) holds.  

(ii) We have
\deqs{
\valj{\tilde{X}\le\tilde{t}}
&=
\valj{\tit\subseteq\tilde{X}}
=
\valj{(\forall x\in\tit) x\in\tilde{X}}
=
\Inf_{q\in\dom(\tit)}\tit(q)\Thenj\valj{q\in\tX}\\
&=
\Inf_{q\in\Q}\tit(\cq)\Then\valj{\cq\in\tX}
=
\Inf_{q\in\Q:t\le q}\valj{\cq\in\tX}
=
\Inf_{q\in\Q:t\le q}\tX(q)
=
E^{X}(t)\\
&=
E^{X}((-\infty,t]).
}
Consequently, (iii) holds.

The rest of the assertions follow similarly.
\eProof 

For any self-adjoint operators $X,Y\in\SA(\cM)$,write $X\les Y$ if 
$E^{Y}(\la)\le E^{X}(\la)$ for all $\la\in\R$.  The relation is called the 
{\em spectral order}.  
This order was originally introduced by Olson \cite{Ols71} for bounded operators; 
for recent results for unbounded operators, see \cite{PS12}.
With the spectral order, the set $\SA(\cM)$ is a conditionally complete
lattice, but it is not a vector lattice, in contrast to the fact that the usual linear order 
$\le$ of self-adjoint operators in $\cM$ is a lattice if and only if $\cM$ is abelian.  
The following facts about the spectral order are known \cite{Ols71,PS12}:
\begin{enumerate}[(i)]\itemsep=0in
\item The spectral order coincides with the usual linear order on projections
and mutually commuting operators.
\item For any $0\le X,Y\in\SA(\cM)$, we have $X\les Y$ if and only if $X^{n}\le Y^{n}$
for all $n\in\N$.
\eenum

The following theorem suggests that the spectral order is coherent with the
natural order defined for internal reals in quantum set theory.

\bTheorem
For any $X,Y\in\SA(\cM)$ and $j=\rS,\rC,\rR$, 
the relation  $X\les Y$ holds if and only $\vval{\tX \le \tY}_j=1$.

\eTheorem

\begin{proof}
From the Takeuti correspondence, we have
\deqs{
\valj{\tX \le \tY}&=\valj{(\forall r\in \tY)r\in \tX}
=\dInf_{r\in\dom(\tY)}(\tY(r)\Thenj \valj{r\in\tX})\\
&=\dInf_{r\in\Q}(E^{Y}(r)\Thenj E^{X}(r)).
}
Thus, the assertion follows from the fact that $E^{Y}(r)\le E^{X}(r)$
if and only if $E^{Y}(r)\Thenj E^{X}(r)=1$ by a minimum implicative
condition, (E).
\end{proof}

\subsection{Experimental meanings for quantum observables}
To explore the experimental meaning of the truth values $\valj{\tX\le \tY}$ 
for $j=\rS,\rC,\rR$ in quantum mechanics, in what follows, we shall confine 
our attention to the case where $\cM$ is the algebra $\rB(\cH)$ of operators 
on a finite-dimensional Hilbert space $\cH$.  

\begin{figure}[H]
  \centering
\includegraphics[width=0.8\textwidth]{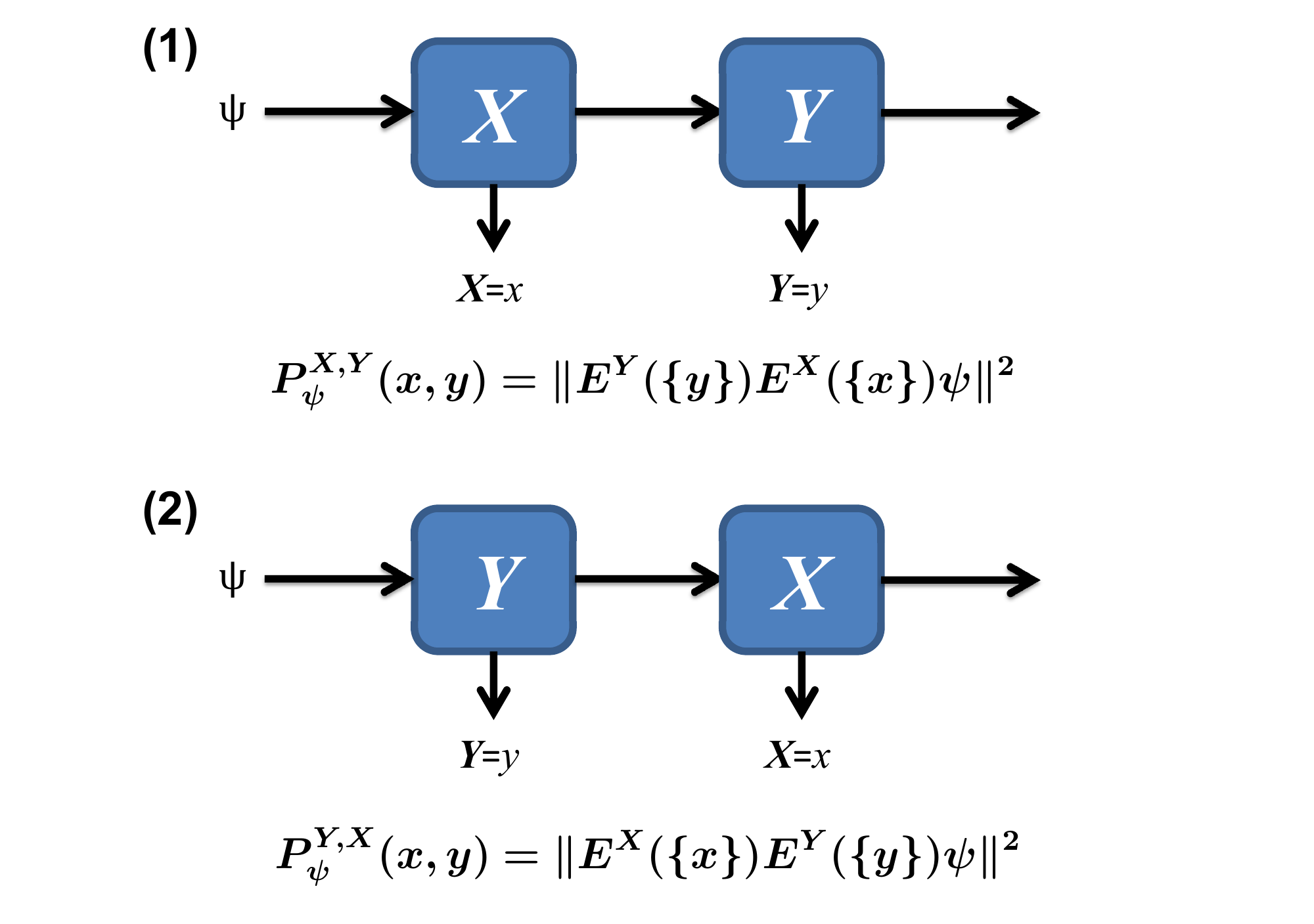}
\caption{Successive projective measurements. (1) 
The observable $X$ is projectively measured with outcome $X=x$ 
in the state $\ps$, and immediately afterward,
the observable $Y$ is measured with outcome $Y=y$. The pair
of outcomes $(X,Y)=(x,y)$ is obtained with probability 
$P^{Y,X}_{\psi}(x,y)$. (2) 
The observable $Y$ is projectively measured with outcome $Y=y$ 
in the state $\ps$, and immediately afterward,
the observable $X$ is measured with outcome $X=x$.
The pair of outcomes $(X,Y)=(x,y)$ is obtained with probability 
$P^{X,Y}_{\psi}(x,y)$.
 }
 \label{fig:one}
\end{figure}

Let $\bS$ be a quantum system described by a finite-dimensional Hilbert space $\cH$.
Following the axioms of quantum mechanics \cite{vN32E},
an {\em observable} of $\bS$ is represented by a self-adjoint operator 
$X$ on $\cH$, and a {\em state} of $\bS$ is represented by a unit vector $\ps$
in $\cH$.  
According to the Born statistical formula \cite{vN32E}, if an observable $X$ 
is measured in a state $\ps$, the probability $P^{X}_{\ps}(x)$ of obtaining the outcome $X=x$ for a real number $x\in\R$
is given by
\deq{P^{X}_{\ps}(x)=\|E^{X}(\{x\})\ps\|^2.}   
Furthermore, according to the von Neumann-L\"{u}ders projection postulate \cite{Lud51}, 
if an observable $X$ is {\em projectively} measured in a state $\ps$ 
and an observable $Y$ is measured immediately afterward, the joint probability
$P^{X,Y}_{\psi}(x,y)$  of the outcomes
 $X=x$ and $Y=y$ for a pair $(x,y)\in\R^2$ of real numbers is given by
  \deq{P^{X,Y}_{\psi}(x,y) =\|E^{Y}(\{y\})E^{X}(\{x\})\ps\|^2 .} 
On the other hand,
if the observable $Y$ is {\em projectively} measured first in the state $\ps$ 
and an observable $X$ is measured immediately afterward, the joint probability
$P^{Y,X}_{\psi}(x,y)$  of the outcomes
 $X=x$ and $Y=y$ for the pair $(x,y)\in\R^2$ of real numbers is given by
 \deq{P^{Y,X}_{\psi}(x,y) =\|E^{X}(\{x\})E^{Y}(\{y\})\ps\|^2 .} 
Then obviously, we have 
$P^{X,Y}_{\psi}(x,y)=P^{Y,X}_{\psi}(x,y)$ for all $(x,y)\in\R^2$
if $XY=YX$, but not necessarily the case if $XY\ne YX$.

We shall write
\deq{
P^{X,Y}_{\psi}(X\le Y)&=\sum_{(x,y):x\le y}P^{X,Y}_{\psi}(x,y),\\
P^{Y,X}_{\psi}(X\le Y)&=\sum_{(x,y):x\le y}P^{Y,X}_{\psi}(x,y).
}

Then we have the following.

\begin{Theorem}\label{th:order-2}
Let $X$ and $Y$ be observables on a finite-dimensional Hilbert space $\cH$
and $\psi$ be a state in $\cH$.   Then we have the following.
\begin{enumerate}[\textrm (i)]\itemsep=0in
\item $\psi\in\cR(\valS{\tX\le \tY})$ if and only if $P^{Y,X}_{\psi}(X\le Y)=1$.
\item $\psi\in\cR(\valC{\tX\le \tY})$ if and only if 
$P^{X,Y}_{\psi}(X\le Y)=1$.
\item $\psi\in\cR(\valR{\tX\le \tY})$ if and only if 
$P^{Y,X}_{\psi}(X\le Y)=P^{X,Y}_{\psi}(X\le Y)=1$.
\eenum
\end{Theorem}
\bproof
Let 
$X=\sum_{k=1}^{n}x_k E^{X}(\{x_k\})$ and 
$Y=\sum_{k=1}^{m}y_k E^{X}(\{y_k\})$ 
be the spectral decompositions of $X$ and $Y$, where $x_1<\cdots<x_n$ and $y_1<\cdots<y_m$.
Then we have 
\deqs{
E^{X}(\la)&=\sum_{k:x_k\le \la} E^{X}(\{x_k\}),\\
E^{Y}(\la)&=\sum_{k:y_k\le \la} E^{Y}(\{y_k\})
}
for all $\la\in\R$.
Let $\psi\in\cR(\valS{\tX\le\tY})$.  
From Proposition \ref{th:order} (i), we have $\ps\in\cR(\tY(\ckr)\ThenS\tX(\ckr))$
for all $r\in\Q$.
From Theorem \ref{th:equivalence-1} (i),  
we have $\tX(\ckr)\p\tY(\ckr)\ps=0$ for all $r\in\Q$.
By the relations $E^X(\la)=\Inf_{r\in\Q:\la\le r}\tX(\ck{r})$ and
$E^Y(\la)=\Inf_{r\in\Q:\la\le r}\tY(\ck{r})$,  we have
$E^{X}(\la)^\perp E^{Y}(\la)\psi=0$ for any $\la\in\R$.
Now, we shall show the relation
\deql{r}{
E^{X}(\la)^\perp E^{Y}(\{\la\})\psi= 0
}
for any $\la\in\R$.
If $\la$ is not an eigenvalue of $Y$, we have $E^{Y}(\{\la\})=0$,
and relation \eq{r} follows.  Suppose $\la=y_k$.  If $k=1$, then 
$E^{Y}(\la)=E^{Y}(\{\la\})$, and hence relation \eq{r} follows.
By induction, we assume $E^{X}(y_j)^\perp E^{Y}(\{y_j\})\psi= 0$ for all $j<k$.
Since $E^{X}(y_k)^{\perp}E^{X}(y_j)^\perp=E^{X}(y_k)^{\perp}$, we have
$E^{X}(y_k)^\perp E^{Y}(\{y_j\})\psi= 0$ for all $j<k$.
Thus, we have $E^{X}(y_k)^\perp E^{Y}(y_{k-1})\psi=\sum_{j=1}^{k-1}
E^{X}(y_k)^\perp E^{Y}(\{y_j\})\psi= 0$.
It follows that 
\deqs{
E^{X}(\la)^\perp E^{Y}(\{\la\})\psi
&=E^{X}(y_{k})^\perp E^{Y}(\{y_{k}\})\psi\\
&=E^{X}(y_{k})^\perp[E^{Y}(y_{k})-E^{Y}(y_{k-1})]\psi\\
&=E^{X}(y_{k})^\perp E^{Y}(y_{k})\psi\\
&=0.}
Thus, relation \eq{r} holds for any $\la\in\R$.
If $x>y$, then
\deqs{
E^{X}(y)\p=\sum_{x'\in\R:x'>y}E^{X}(\{x'\})\ge E^{X}(\{x\}),
}
and hence
\deqs{P^{Y,X}_{\psi}(x,y)=\|E^{X}(\{x\}) E^{Y}(\{y\})\psi\|^2
\le \|E^{X}(y)\p E^{Y}(\{y\})\psi\|^2
= 0.
}
Thus,
\deqs{
P^{Y,X}_{\psi}(X\le Y)&=\sum_{(x,y)\in\R^{2}:x\le y}P^{Y,X}_{\psi}(x,y)
=1-\sum_{(x,y)\in\R^{2}:x>y}P^{Y,X}_{\psi}(x,y)=1.
}
Conversely, suppose $P^{Y,X}_{\psi}(X\le Y)=1$.  
Then we have $E^{X}(\{x\})E^{Y}(\{y\})\psi=0$ for all
$x>y$.
Consequently,
\deqs{
E^{X}(\la)^{\perp}E^{Y}(\la)\psi
&=\Big(\sum_{x\in\R:\la<x}E^{X}(\{x\})\Big)
   \Big(\sum_{y\in\R:y\le\la}E^{Y}(\{y\})\Big)\ps\\
&=\sum_{(x,y)\in\R^{2}:y\le\la<x}E^{X}(\{x\})E^{Y}(\{y\})\ps
=0} for every $\la\in\R$.
Thus,  from Theorem  \ref{th:equivalence-1} (i),
$\ps\in\cR(\tY(\ckr)\ThenS\tX(\ckr))$ for all $r\in\Q$,
so 
\deqs{\ps\in \bigcap_{r\in\Q}\cR(\tY(\ckr)\ThenS\tX(\ckr))
=\cR\Big(\Inf_{r\in\Q}(\tY(\ckr)\ThenS\tX(\ckr))\Big)
=\cR(\valS{\tX\le\tY}).
}
Therefore, assertion (i) follows.

The rest of the assertions follow similarly.
\eproof

Note that  $P^{Y,X}_{\psi}(X\le Y)=1$ means that 
the outcome of the 
$X$-measurement is less than or equal to the outcome of the $Y$-measurement 
 with probability 1 in a successive $(Y,X)$-measurement.
Similarly, $P^{X,Y}_{\psi}(X\le Y)=1$ means that
the outcome of the $Y$-measurement is greater than or equal to the outcome
of the $X$ measurement
in a successive $(X,Y)$-measurement with probability 1.

Thus, we have completely characterized the experimental meanings of
the order relations for the internal reals derived from each choice of the conditional operation
in terms of joint probability obtained by the successive projective measurements of two arbitrary observables.
It is clearly shown that those operations reveal their individual features 
with symmetry between the Sasaki conditional and the contrapositive Sasaki conditionals
with respect to the order of measurement, 
in contrast to the prevailing view that favors the Sasaki conditional. 
Moreover, the relevance conditional has meaning as a conjunction of both the Sasaki conditional and the contrapositive
Sasaki conditional, and achieves self-symmetry, which would be preferable to the other two in some contexts.

\subsection{Examples}
Continuing the setting of the preceding subsection, 
we consider the case where $\cM=\cB(\cH)$ for a finite dimensional Hilbert space $\cH$.
First, we consider the case where $X,Y\in\SA(\cM)$ are projections $X=P\in\cQ(\cH)$ and $Y=Q\in\cQ(\cH)$.

From \cite[Theorem 3.3, Proposition 3.7]{17A2} we have the following relations
for $P\Thenj Q$ with $j=S,C,R$.
\benum
\item $P\ThenS Q=Q\p\ThenC P\p$.
\item $P\ThenC Q=Q\p\ThenS P\p$.
\item $P\ThenR Q=P\ThenS Q\And P\ThenC Q$. 
\item $P\ThenS Q=(P\And Q)\Or (P\And Q\p) \Or (P\p\And Q\p)\Or(Q\And \com(P, Q)\p)$.
\item $P\ThenC Q=(P\And Q)\Or (P\And Q\p) \Or (P\p\And Q\p)\Or(P\p\And \com(P, Q)\p)$.
\item $P\ThenR Q=(P\And Q)\Or (P\And Q\p) \Or (P\p\And Q\p)$.
\item $P\ThenR Q=(P\ThenS Q)\And \com(P,Q)=(P\ThenC Q)\And \com(P,Q)\le \com(P,Q)$.
\eenum

In this case, 
\deq{E^{P}(r)=
\begin{cases}
0 & (\la < 0)\\
P\p & (0 \le \la <1)\\
I & (1<\la),
\end{cases}
\quad
E^{Q}(r)=
\begin{cases}
0 & (\la < 0)\\
Q\p & (0 \le \la <1)\\
I & (1\le\la).
\end{cases}
}
By relation (i) we have
$E^{Q}(r)\ThenS E^{P}(r)=
Q\p\ThenS P\p=P\ThenC Q$
for $0\le \la< 1$ and $E^{Q}(r)\ThenS E^{P}(r)=1$ otherwise.
Similarly, we have $E^{Q}(r)\ThenC E^{P}(r)=P\ThenS Q$ and
$E^{Q}(r)\ThenR E^{P}(r)=P\ThenR Q$ for $0\le \la<1$ 
and $E^{Q}(r)\ThenC E^{P}(r)=E^{Q}(r)\ThenR E^{P}(r)=1$ otherwise.
It follows that 
\deqed{
\valS{\tP\le \tQ}&=P\ThenC Q,\\
\valC{\tP\le \tQ}&=P\ThenS Q,\\
\valR{\tP\le \tQ}&=P\ThenR Q.
}

Suppose $\dim(\cH)=2$ and that $P,Q$ are nontrivial, i.e., $0<P,Q<I$.
Then, $\com(P,Q)=0$.
So, from relations (iv)--(vii), we have  
\deqed{
\valS{\tP\le \tQ}&=P\p,\\
\valC{\tP\le \tQ}&=Q,\\
\valR{\tP\le \tQ}&=0.
}
Thus, from Theorem \ref{th:order-2} we obtain.
\bProposition
Suppose $\cM=\cB(\cH)$ and $P,Q\in\cQ(\cH)$.
Then, we have the following statements. 
\benum
\item $P^{P,Q}_{\psi}(P\le Q)=1$ if and only if $\ps\in\cR(P\ThenC Q)$.\\
\item $P^{Q,P}_{\psi}(P\le Q)=1$ if and only if $\ps\in\cR(P\ThenS Q)$.\\
\item $P^{Q,P}_{\psi}(P\le Q)=P^{P,Q}_{\psi}(P\le Q)=1$ if and only if $\ps\in\cR(P\ThenS Q)$.
\eenum
Furthermore, if $\dim(\cH)=2$, then we have the following statements.
\begin{enumerate}
\item[{\rm (iv)}] $P^{P,Q}_{\psi}(P\le Q)=1$ if and only if $\ps\in\cR(P)\p$.
\item[{\rm (v)}]  $P^{Q,P}_{\psi}(P\le Q)=1$ if and only if $\ps\in\cR(Q)$.
\item[{\rm (vi)}]  No state vector satisfies $P^{Q,P}_{\psi}(P\le Q)=P^{P,Q}_{\psi}(P\le Q)=1$.
\end{enumerate}
\eProposition
The above theorem reveals the mutually symmetric roles of the Sasaki conditional and the contrapositive Sasaki conditional, whereas
the Sasaki conditional has often been considered to be superior to
the contrapositive Sasaki conditional.  

In quantum cognition, the successive projective measurements depicted 
in  Figure \ref{fig:one} has been 
proposed to model the question order effect known for opinion poll data
\cite{WB13,WSSB14}.
For instance, consider two questions $A,B$ in an opinion poll such as
\begin{quotation}
$A:$ ``Do you generally think Bill Clinton is honest and trustworthy?'' 

$B:$ ``Do you generally think Al Gore is honest and trustworthy?'' 
\end{quotation}
Then, we denote by $P(AaBb)$ $(a,b\in\{{\rm y(es)}, {\rm n(o)}\})$ 
the joint probabilities of answering $a$ to question $A$ and then answering $b$ to
question $B$ and by $P(BbAa)$ the joint probabilities of answers to questions
$A,B$ asking in the opposite order.
A projective measurement model is given by a quadruple $(\cH,P,Q,\psi)$
consisting of a Hilbert space $\cH$, two projections $P,Q$ on $\cH$, and
a unit vector $\ps$ in $\cH$ to model the joint probabilities $P(AaBb)$
and $P(BbAa)$ as the outcomes of successive projective measurements
of the observable $P,Q\in\cQ(\cH)$ so that
\deqed{
P(AaBb)=P_{\ps}^{P,Q}(x,y)=\|E^{Q}(\{y\})E^{P}(\{x\})\ps\|^2,\\
P(BbAa)=P_{\ps}^{Q,P}(x,y)=\|E^{P}(\{x\})E^{Q}(\{y\})\ps\|^2,
}
where we encode $x, y=1, 0$ corresponding to $a, b={\rm y}, {\rm n}$.  
In this model, $P^{P,Q}_{\psi}(P\le Q)=1$ means that the model with
the state $\ps$ predicts that question $B$ certainly receives the favorable or
even answer compared with question $A$, provided that question $B$ is asked second, 
whereas $P^{Q,P}_{\psi}(P\le Q)=1$ means that the model with
the state $\ps$ predicts that question $B$ certainly receives favorable 
answer to question $A$ provided that question $B$ is asked first.
It would be interesting that if $\dim(\cH)=2$, statements (iv)--(vi) holds. 

For further study in the relation between other models of the
question order effect \cite{OK21JMP,23A1}
and the experimental meaning of the truth values 
$\valj{\tP\le\tQ}$ for $j=\rS,\rC,\rR$ 
with quantum computer realizations will be discussed elsewhere.

\section{Conclusion}
In quantum logic, there are at least three candidates for conditional operation,
called the Sasaki conditional, the contrapositive Sasaki conditional,
and the relevance conditional.  In this study, we have attempted to
develop a quantum set theory based on quantum logics with those three 
conditionals, each of which defines a different quantum logical truth
value assignment.  In a previous study \cite{21QTD}, we showed 
that those three models 
satisfy the transfer principle of the same form to determine the quantum 
logical truth values of theorems of the ZFC set theory.
In this study, we have explored the structures of the models' internal reals.
We have shown that the truth values of their equality 
are the same for those models.
Interestingly, however, we have revealed  that the order relation between 
them significantly depends on the underlying conditionals.  
In particular, we have completely characterized the experimental meanings of 
those order relations in terms of joint probability obtained by the successive
projective measurements of two arbitrary observables.
Those characterizations clearly show their individual features
with symmetry between the Sasaki and the contrapositive Sasaki conditionals, in contrast to the majority view that favors the Sasaki conditional. 
Furthermore, the relevance conditional has meaning as a conjunction of both the Sasaki conditional and the contrapositive Sasaki conditional, and as a result achieves self-symmetry.

Our findings reveal that quantum set theory yields empirically testable predictions concerning state-dependent binary relations between quantum observables, 
emerging from well-known relations, such as Olson's spectral order relation, originally formulated state-independently,
thereby considerably extending the standard probabilistic interpretation of quantum mechanics.

\appendix
\section*{Appendix}
\renewcommand{\thesection}{\Alph{section}}
\section{Quantization of logical operations}
\subsection{Quantizations of conditional}
Let $\cQ$ be a complete orthomodular lattice, called a logic for simplicity.  
A binary ortholattice polynomial $\Then$ on a logic $\cQ$ is called a {\em quantized conditional} iff it satisfies 
 \begin{itemize}
 \item[(LB)] If $P\commutes Q$, then 
$P\Then Q=P^{\perp}\Or Q$ for all $P,Q\in\cQ$.
\end{itemize} 

The Kotas theorem \cite{Kot67} concludes the following.
\bTheorem\label{th:Kotas-3}
{There exist exactly 6 two-variable ortholattice-polynomials $P\Thenj Q$ for $j=0,
\ldots,5$
satisfying (LB), given as follows.}
\benum
\item[{\rm (0)}] 
$P\Then_0 Q=b_n(P,Q)$.
\item[{\em (1)}] 
$P\Then_1 Q=b_n(P,Q)\Or (P\And \com(P,Q)\p)$.
\item[{\em (2)}] 
$P\Then_2 Q=b_n(P,Q)\Or (Q\And \com(P,Q)\p)$.
\item[{\em (3)}] 
$P\Then_3 Q=b_n(P,Q)\Or (P\p\And \com(P,Q)\p)$.
\item[{\em (4)}] 
$P\Then_4 Q=b_n(P,Q)\Or (Q\p \And \com(P,Q)\p)$.
\item[{\em (5)}] 
$P\Then_5 Q=b_n(P,Q)\Or \com(P,Q)\p$.
\end{enumerate}
In the above, $b_n(P,Q)$ is the disjunctive normal form of the classical
conditional $P\p\Or Q$, \ie,
\[
b_n(P,Q)=(P\p\And Q\p)\Or(P\p\And Q)\Or(P\And Q).
\]
\eTheorem

For $j=0,\ldots,5$, the above polynomials $P\Thenj Q$ are explicitly expressed as follows.
\benum\item[{\rm (0)}] 
 $P\Then_0 Q=(P\p\And Q\p)\Or(P\p\And Q)\Or(P\And Q)$.
\item[{\rm (1)}] 
 $P\Then_1 Q=(P\p\And Q\p)\Or(P\p \And Q)\Or(P\And(P\p\Or Q))$.
\item[{\rm (2)}] 
 $P\Then_2 Q=(P\p\And Q\p)\Or Q$.
\item[{\rm (3)}] 
 $P\Then_3 Q=P\p \Or(P\And Q)$.
\item[{\rm (4)}] 
 $P\Then_4 Q=((P\p\Or Q)\And Q\p)\Or(P\p \And Q)\Or(P\And Q)$.
\item[{\rm (5)}] 
 $P\Then_5 Q=P\p\Or Q$.
\end{enumerate}

We call
$\Then_0=\ThenR$ the {\em minimum conditional} or {\em relevance conditional} \cite{Geo79},
$\Then_2=\ThenC$ the {\em contrapositive Sasaki conditional},
$\Then_3=\ThenS$ the {\em Sasaki conditional}  \cite{Sas54,Fin69}, and 
$\Then_5$  the {\em classical conditional}.  

\subsection{Quantizations of conjunction}
A binary ortholattice operation $\ast$  on a logic $\cQ$ is called a {\em quantized conjunction} iff it satisfies the following:
\benum\item[{\rm (GC)}] If $P\commutes Q$, then $P\ast Q=P\And Q$.
\end{enumerate}

In Boolean logic, the conditional and conjunction are associated by
the relation $P\And Q=(P\Then Q^{\perp})\p$, and this relation plays
an essential role in the duality between bounded universal quantification
$(\forall x\in u)\ph(x)$ and bounded existential quantification
$(\exists x\in u)\ph(x)$ \cite{21QTD}.

For any quantized conditional $\Then$, we call the operation $*$ defined by
\deql{DC}{
P*Q=(P\Then Q\p)\p
}
for all $P,Q\in\cQ$
the {\em dual conjunction} of $\Then$.

For any $j=0,\ldots,5$, denote by $\ast_j$ the dual conjunction of the quantized
conditional $\Then_j$. Then we have
\benum
\item[{\rm (0)}] 
$P\ast_0 Q=(P\And Q)\Or \com(P,Q)\p$.
\item[{\rm (1)}] 
$P\ast_1 Q=(P\And Q)\Or (P\p\And \com(P,Q)\p)$.
\item[{\rm (2)}] 
$P\ast_2 Q=(P\And Q)\Or (Q\And \com(P,Q)\p)$.
\item[{\rm (3)}] 
$P\ast_3 Q=(P\And Q)\Or (P\And \com(P,Q)\p)$.
\item[{\rm (4)}] 
$P\ast_4 Q=(P\And Q)\Or (Q\p \com(P,Q)\p)$.
\item[{\rm (5)}] 
$P\ast_5 Q=P\And Q$.
\eenum

We call $\ast_5$ the {\em classical conjunction}, and $\ast_3$ the {\em Sasaki conjunction}.

We have the following.
\bProposition\label{th:dual-conjunction}
A binary operation $*$ on a logic $\cQ$ is a quantized conjunction if and only if
it is the dual conjunction of a quantized conditional $\Then$ on $\cQ$.
\eProposition

\section{Quantum set theory}
\subsection{Orthomodular-valued interpretations}
\label{se:interpretations}
Let $\cLL(\in)$ be the language of first-order theory with equality 
consisting of the negation symbol $\Not$; connectives $\And, \Or, 
\Implies$; binary relation symbols $=,\in$; bounded 
quantifier symbols $\forall x\in y$, $\exists x \in y$; unbounded
quantifier symbols $\forall x, \exists x$; and no constant symbols.
For any class $U$, the language $\cLL(\in,U)$ is the one obtained 
by adding a name for each element of $U$.

To each statement $\ph$ of $\cLL(\in,U)$, the satisfaction
relation
$\bracket{U,\in} \models \ph$ is defined by the following recursive rules:
\benum
\item[(i)]$ \bracket{U,\in} \models \Not \ph \iff  \bracket{U,\in}
\models
\ph 
\mbox{ does not hold}$. 
\item[(ii)] $ \bracket{U,\in} \models \ph_1 \And \ph_2 
\iff \bracket{U,\in}
\models \ph_1 
\mbox{ and } \bracket{U,\in} \models \ph_2$.
\item[(iii)] $ \bracket{U,\in} \models \ph_1 \Or \ph_2 
\iff \bracket{U,\in}
\models \ph_1 
\mbox{ or } \bracket{U,\in} \models \ph_2$.
\item[(iv)] $ \bracket{U,\in} \models \ph_1 \Then \ph_2 
\iff \mb{ if }\bracket{U,\in}\models \ph_1 
\mbox{, then } \bracket{U,\in} \models \ph_2$.
\item[(v)] $ \bracket{U,\in} \models  (\forall x\in u)\,\ph(x) \iff
\bracket{U,\in} \models \ph(u') \mbox{ for all } u' \in u$.
\item[(vi)] $ \bracket{U,\in} \models  (\exists x\in u)\,\ph(x) \iff
\mbox{there exists $u' \in u$ such that $\bracket{U,\in} \models \ph(u')$.}$
\item[(vii)] $ \bracket{U,\in} \models  (\forall x)\,\ph(x) \iff
\bracket{U,\in} \models \ph(u) \mbox{ for all } u \in U.$
\item[(viii)] $ \bracket{U,\in} \models  (\exists x)\,\ph(x) \iff
\mbox{ there exists }u\in U\mbox{ such that }
\bracket{U,\in} \models \ph(u) $.
\item[(ix)] $ \bracket{U,\in} \models u = v
\iff u = v$. 
\item[(x)] $ \bracket{U,\in} \models u\in v
\iff u\in v$.
\end{enumerate}
Our assumption that $\V$ is the universe of ZFC
means that if $\mb{ZFC}\vdash \ph(x_1,\ldots,x_n)$, then 
$\bracket{\V,\in}\models \ph(u_1,\ldots,u_n)$ for 
any formula $\ph(x_1,\ldots,x_n)$ of $\cLL(\in)$ 
provable in ZFC and for any $u_1,\ldots,u_n\in \V$.

\renewcommand{\Thenj}{\rightarrow}

\sloppy
Let $\cQ$ be a logic.
Let $\VQ$ be the $\cQ$-valued universe of quantum set theory.
Denote by $\cSS(\cQ)$ the class of statements in $\cLL(\in,\VQ)$.
A {\em $\cQ$-valued interpretation} of $\cLL(\in,\VQ)$ is a mapping 
$\cI{(\Then,\ast)}:\ph\in\cSS(\cQ)\mapsto \valQ{\ph}\in\cQ$ determined with a pair 
$(\Then,\ast)$ of binary operations on $\cQ$ by 
the following rules, (R1)--(R10),
recursive on the rank of elements of $\VQ$ and the complexity of formulas.

\benum
\item[(R1)]  
$ \valQ{\Not\ph} = \valQ{\ph}^{\perp}$.
\item[(R2)]  $ \valQ{\ph_1\And\ph_2} 
= \valQ{\ph_{1}} \And \valQ{\ph_{2}}$.
\item[(R3)]  $ \valQ{\ph_1\Or\ph_2} 
= \valQ{\ph_{1}} \Or \valQ{\ph_{2}}$.
\item[(R4)]  $ \valQ{\ph_1\rightarrow\ph_2} 
= \valQ{\ph_{1}} \Then \valQ{\ph_{2}}$.
\item[(R5)]  $ \valQ{(\forall x\in u)\, {\ph}(x)} 
= \dInf_{u'\in \dom(u)}
(u(u') \Thenj \valQ{\ph(u')})$.
\item[(R6)]  $ \valQ{(\exists x\in u)\, {\ph}(x)} 
= \dSup_{u'\in \dom(u)}
(u(u') \ast \valQ{\ph(u')})$.
\item[(R7)]  $ \valQ{(\forall x)\, {\ph}(x)} 
= \dInf_{u\in \VQ}\valQ{\ph(u)}$.
\item[(R8)]  $ \valQ{(\exists x)\, {\ph}(x)} 
= \dSup_{u\in \VQ}\valQ{\ph(u)}$.
\item[(R9)] $\valQ{u = v}
= \dInf_{u' \in  \dom(u)}(u(u') \Thenj
\valQ{u'  \in v})
\And \dInf_{v' \in   \dom(v)}(v(v') 
\Thenj \valQ{v'  \in u})$.
\item[(R10)] $ \valQ{u \in v} 
= \dSup_{v' \in \dom(v)} (v(v')* \valQ{v' = u})$.
\eenum

For a sublogic $\cR$ of a logic $\cQ$ with a $\cQ$-valued interpretation $\cI(\Then,\ast)$,
we denote by $\val{\ph}_{\cR}$ the $\cR$-valued
truth value of a statement $\ph\in\cSS(\cR)$ 
determined by the $\cR$-valued interpretation $\cI(\Then_{\cR},\ast_{\cR})$,
where $\Then_{\cR}$ and $\ast_{\cR}$ are the restrictions 
of $\Then$ and $\ast$ to $\cR$.

A formula in $\cLL(\in)$ is called a {\em
$\De_{0}$-formula}  iff it has no unbounded quantifiers
$\forall x$ or $\exists x$.
The following theorem holds \cite{21QTD}.

\sloppy
\bTheorem[$\De_{0}$-Absoluteness Principle]
\label{th:Absoluteness-A}
\sloppy  
Let $\cR$ be a sublogic of a logic $\cQ$ with a $\cQ$-valued interpretation 
$\cI(\Then,\ast)$ of $\cLL(\in,\VQ)$.
For any $\De_{0}$-formula 
${\ph} (x_{1},{\ldots}, x_{n}) \in \cLL(\in)$ and $u_{1},{\ldots}, u_{n}\in V^{(\cR)}$, 
we have
\[
\val{\ph(u_{1},\ldots,u_{n})}_{\cR}=
\val{\ph(u_{1},\ldots,u_{n})}_{\cQ}.
\]
\eTheorem

Henceforth, for any $\De_{0}$-formula 
${\ph} (x_{1},{\ldots}, x_{n}) \in\cLL(\in)$
and $u_1,\ldots,u_n\in\VQ$,
we abbreviate $\val{\ph(u_{1},\ldots,u_{n})}=
\valQ{\ph(u_{1},\ldots,u_{n})}$.

\subsection{Transfer principle}
\label{se:ZFC}\label{se:TPQ}
In this section, we investigate the transfer principle for general $\cQ$-valued interpretation.
Let $\cI(\Then,\ast)$ be a $\cQ$-valued interpretation.
We denote by $\val{\ph}$ the $\cQ$-valued
truth value of a $\De_0$-statement $\ph$ 
determined by the $\cQ$-valued interpretation $\cI(\Then,\ast)$.
Then the {\em transfer principle} for the $\cQ$-valued interpretation $\cI(\Then,\ast)$
 is formulated as follows.
\medskip

{\bf Transfer Principle.}  
{\em Any $\De_{0}$-formula ${\ph} (x_{1},{\ldots}, x_{n})$ in
$\cLL(\in)$ provable in ZFC satisfies
\deq{
\val{\ph({u}_{1},\ldots,{u}_{n})}\ge
\cuniv(u_{1},\ldots,u_{n})
}
}
 for any $u_1,\ldots,u_n\in\VQ$.
\medskip

A $\cQ$-valued interpretation $\cI(\Then,\ast)$ is called the {\em Takeuti interpretation}
iff $\Then$ is the Sasaki conditional and $\ast$ is the classical conjunction, \ie,
$P\Then Q=P\Then_3 Q=P\p\Or(P\And Q)$ and $P\ast Q=P\ast_5 Q=P\And Q$
for all $P,Q\in\cQ$.
It was shown that if $\cQ$ is the projection lattice of a von Neumann algebra,
then the $\cQ$-valued Takeuti interpretation $\cI(\Then_3,\ast_5)$ satisfies the
transfer principle \cite{07TPQ}.
This result was extended to an arbitrary logic $\cQ$ and arbitrary quantized
conditional $\Then$ on $\cQ$ to show that any $\cQ$-valued interpretation
$\cI(\Then,\ast_5)$ satisfies the transfer principle \cite{17A2}.
In our previous work \cite{21QTD}, we found  all the interpretations that satisfy
the transfer principle.  

In order to eliminate uninteresting interpretations from our consideration,
we call a $\cQ$-valued interpretation $\cI(\Then,\ast)$ {\em non-trivial}
iff for any $P\in\cQ$ there exist a $\De_0$-formula $\ph(x_1,\ldots,x_n)\in\cLL(\in)$
and $u_1,\ldots,u_n\in\VQ$ such that $\cm(u_1,\ldots,u_n)=1$ and 
$\val{\ph(u_1,\ldots,u_n)}=P$.
Simple sufficient conditions for non-triviality are given as follows.
\bProposition
If a $\cQ$-valued interpretation $\cI(\Then,\ast)$ satisfies
\benum
\item[{\rm (i)}] $P\Then 0=P\p$ \  for all $P\in\cQ$, or   
\item[{\rm (ii)}] $P\ast 1=P$ \   for all $P\in\cQ$, 
\eenum
then  $\cI(\Then,\ast)$ is non-trivial.
\eProposition

In what follows, we introduce the connective $\iff$ in the language
$\cLL(\in\VQ)$ as an abbreviation for 
$\ph\iff \ps:=(\ph\And \ps)\Or(\Not \ph\And \Not \ps)$
for any $\ph,\ps\in\cLL(\in,\VQ)$
and the corresponding operation $\iff$ on $\cQ$ by
$P\iff Q:=(P\And Q)\Or(P\p \And Q\p)$ for all $P,Q\in\cQ$.

A $\cQ$-valued interpretation $\cI(\Then,\ast)$ is called {\em normal} iff
$\Then$ is a quantized conditional and $\ast$ is a quantized conjunction.
It is easy to see that all normal interpretations are non-trivial.
 
In what follows,
suppose that for any $\ph\in\cLL(\in,\VQ)$, the truth value $\val{\ph}\in\cQ$
is assigned by a fixed but arbitrary normal $\cQ$-valued interpretation
$\cI(\Then,\ast)$.

The following theorem is known as the fundamental theorem of 
Boolean-valued models of set theory.
\bTheorem\label{th:BTP-1}
If $\cQ$ is a Boolean logic, all the normal interpretations define
the unique $\cQ$-valued interpretation, and it satisfies the following statements.
\benum
\item[{\rm (i)}] $\val{(\exists x\in u)\ph(x)}
=\val{(\exists x)(x\in u\And \ph(u))}$ for every formula $\ph(x)$ in 
$\cLL(\in,\VQ)$ with one free variable $x$ and $u\in\VQ$.
\item[{\rm (ii)}]  $\val{(\forall x\in u)\ph(x)}
=\val{(\forall x)[\Not(x\in u)\Or\ph(x)]}$
for every formula $\ph(x)$ in 
$\cLL(\in,\VQ)$ with one free variable $x$ and $u\in\VQ$.
\item[{\rm (iii)}] $\val{\ph}=1$ for any statement in $\cLL(\in,\VQ)$ provable in ZFC.
\eenum
\eTheorem

Denote by ${\bf 2}$ the sublogic ${\bf 2}=\{0,1\}$ in any logic $\cQ$.
We have the following.
\bTheorem[$\De_0$-Elementary Equivalence Principle]
\label{th:2.3.2-A}
\sloppy
Let ${\ph} (x_{1},{\ldots}, x_{n}) $ be a 
$\De_{0}$-for\-mu\-la  of $\cLL(\in)$.
For any $u_{1},{\ldots}, u_{n}\in V$,
we have
\[
\bracket{\V,\in}\models {\ph}(u_{1},{\ldots},u_{n})
\quad\mbox{if and only if}\quad
\val{\ph(\check{u}_{1},\ldots,\check{u}_{n})}=1.
\]
\eTheorem

Let $\cA\subseteq\VQ$.  The {\em logic
generated by $\cA$}, denoted by $\cQ(\cA)$, is  defined by 
\deq{
\cQ(\cA)=L(\cA)^{!!}.
}
For $u_1,\ldots,u_n\in\VQ$, we write
$\cQ(u_1,\ldots,u_n)=\cQ(\{u_1,\ldots,u_n\})$.

 The following theorem characterizes non-trivial $\cQ$-valued interpretations
 that satisfy the transfer principle.
 
\bTheorem\label{th:TP-3}
A non-trivial $\cQ$-valued interpretation $\cI(\Then,\ast)$ satisfies the transfer principle if and only if it is normal.
Non-trivial $\cQ$-valued interpretations 
$\cI(\Then,\ast)$ of $\cLL(\in,\VQ)$
satisfying the transfer principle are unique
if $\cQ$ is a Boolean algebra, and there are exactly 36 such $\cQ$-valued
interpretations $\cI(\Then_j,\ast_k)$ for $j,k=0,\ldots,5$
 if $\cQ$ is not a Boolean algebra.
\eTheorem

\subsection{De Morgan's laws}
 
 Every $\cQ$-valued interpretation $\cI(\Then,\ast)$ with an arbitrary pair $(\Then,\ast)$
 of binary polynomial operations satisfies De Morgan's laws
for conjunction-disjunction connectives and for universal-existential quantifiers
simply according to the duality between supremum and infimum as follows. 
\benum
\item[(M1)] $\val{\Not(\ph_1\And\ph_2)} =\val{\Not \ph_1\Or \Not \ph_2},$
\item[(M2)] $\val{\Not(\ph_1\Or \ph_2)}=\val{\Not \ph_1\And \Not \ph_2},$
\item[(M3)] $\val{\Not(\forall x\,\ph(x))}=\val{\exists x\,(\Not \ph(x))},$
\item[(M4)] $\val{\Not(\exists x\,\ph(x))}=\val{\forall x\,(\Not \ph(x))}.$
 \end{enumerate}
 
 However, De Morgan's laws for bounded quantifiers, 
 \benum
\item[(M5)] $\val{\Not(\forall x\,\ph(x))}=\val{\exists x\,(\Not \ph(x))},$
\item[(M6)] $\val{\Not(\exists x\,\ph(x))}=\val{\forall x\,(\Not \ph(x))},$
\end{enumerate}
are not generally  satisfied, even for normal interpretations, as shown below.
 
A $\cQ$-valued interpretation $\cI(\Then,\ast)$ of $\cLL(\in,\VQ)$ is called
the {\em Takeuti interpretation} iff $\Then=\Then_3$ and $\ast=\ast_5=\And$.
The Takeuti interpretation $\cI(\Then_3,\ast_5)$ was introduced by Takeuti \cite{Ta81} for
the projection lattice $\cQ=\cQ(\cH)$ on a Hilbert space $\cH$.
The present author extended it to the projection lattice $\cQ=\cQ(\cM)$ of a von Neumann
algebra $\cM$ in \cite{07TPQ}, and to a general complete orthomodular
lattice $\cQ$ in \cite{17A2}.
This is only one interpretation of quantum set theory that has been studied 
seriously so far \cite{Yin05,Eva15,16A2,17A1,21BBQ}.

A $\cQ$-valued interpretation $\cI(\Then,\ast)$ of $\cLL(\in,\VQ)$ is said to be
{\em self-dual} iff 
\[
P\ast Q=(P\Then Q\p)\p
\] 
for all $P,Q\in\cQ$.

\bTheorem
A non-trivial $\cQ$-valued interpretation $\cI(\Then,\ast)$ of $\cLL(\in,\VQ)$
satisfies De Morgan's laws if and only if 
it is self-dual.
\eTheorem

Now, we conclude:
\bTheorem
A  $\cQ$-valued interpretation $\cI(\Then,\ast)$ satisfies both the
Transfer
Principle and De Morgan's laws if and only if $\Then$ is a quantized conditional
and $\ast$ is its dual conjunction, namely, the $\cQ$-valued interpretation $\cI(\Then,\ast)$
is normal and self-dual.
\eTheorem

In the main text, we investigate the normal self-dual interpretation  
$\cI(\Then_j, \ast_j)$  for $j=0,2,3$. 
For a normal self-dual $\cQ$-valued interpretation $\cI(\Then,\ast)$ 
of $\cLL(\in,\VQ)$,
we can take the symbols $\Not$, $\And$, $\Then$, $\forall x\in y$, and $\forall x$
as primitive as in the language $\LL(\in)$, and the symbols $\Or$, 
$\exists x\in y$, and $\exists x$ as derived symbols.   

Now, we conclude the following  characterization of
$\cQ$-valued interpretations that satisfy both the transfer principle and
De Morgan's laws.

\bTheorem
Let $\cQ$ be a logic and $(\Then,\ast)$ be a pair of two-variable ortholattice
polynomials. Then the $\cQ$-valued interpretation $\cI(\Then,\ast)$ of 
$\cLL(\in,\VQ)$
satisfying both the transfer principle and De Morgan's laws is unique
if $\cQ$ is a Boolean algebra and exactly six such interpretations, \ie,
$\cI(\Then_j,\ast_j)$ for $j=0,\ldots,5$,
 if $\cQ$ is not Boolean.
\eTheorem

\section*{Acknowledgements}
This work was supported by the RIKEN TRIP Initiative (RIKEN Quantum).
The author acknowledges the support of the Quantinuum--Chubu University Collaboration in 2023--2024.

\section*{Declarations}
\begin{itemize}
\item Funding: This work was supported by JSPS KAKENHI Grant Numbers JP25K07108, JP24H01566,
and JST CREST Grant Number JPMJCR23P4, Japan.
\item Conflict of interest/Competing interests: Not applicable.
\item Consent for publication: Not applicable. 
\item Data availability: Not applicable. 
\item Materials availability: Not applicable. 
\item Code availability: Not applicable.  
\item Author contribution: The author solely conducted the research, analysis, and writing of the manuscript.
\end{itemize}

\end{document}